\documentclass[cp]{copernicus}
\pdfoutput=1
\usepackage{longtable}

\begin{document}%\hack{\sloppy}

\title{Radiative forcings for 28~potential Archean greenhouse gases}

\Author{B.}{Byrne}
\Author{C.}{Goldblatt}

\affil{School of Earth and Ocean Sciences, University of Victoria, Victoria, BC, Canada}

\correspondence{B.~Byrne (bbyrne@uvic.ca)}

\runningtitle{Archean radiative forcings}
\runningauthor{B.~Byrne and C.~Goldblatt}

\received{2 April 2014}
\accepted{22 April 2014}
\published{}

\firstpage{1}

\maketitle

\begin{abstract}
Despite reduced insolation in the late Archean, evidence suggests a~warm climate which was likely sustained by a~stronger greenhouse effect, the so-called Faint Young Sun Problem (FYSP). \chem{CO_2} and \chem{CH_4} are generally thought to be the mainstays of this enhanced greenhouse, though many other gases have been proposed. We present high accuracy radiative forcings for \chem{CO_2}, \chem{CH_4} and 26 other gases, performing the radiative transfer calculations at line-by-line resolution and using HITRAN 2012 line data for background pressures of 0.5, 1, and 2\,\unit{bar} of atmospheric $\mathrm{N_2}$. For \chem{CO_2} to resolve the FYSP alone at 2.8~Gyr BP (80\,{\%} of present solar luminosity), 0.32\,\unit{bar} is needed with 0.5\,\unit{bar} of atmospheric $\mathrm{N_2}$, 0.20\,\unit{bar} with 1\,\unit{bar} of atmospheric $\mathrm{N_2}$, or 0.11\,\unit{bar} with 2\,\unit{bar} of atmospheric $\mathrm{N_2}$. For \chem{CH_4}, we find that near-infrared absorption is much stronger than previously thought, arising from updates to the HITRAN database. \chem{CH_4} radiative forcing peaks at 10.3, 9, or 8.3\,\unit{W\,m^{-2}} for background pressures of 0.5, 1 or 2\,\unit{bar}, likely limiting the utility of \chem{CH_4} for warming the Archean. For the other 26 HITRAN gases, radiative forcings of up to a~few to 10\,\unit{W\,m^{-2}} are obtained from concentrations of 0.1--1~ppmv for many gases. For the 20 strongest gases, we calculate the reduction in radiative forcing due to overlap. We also tabulate the modern sources, sinks, concentrations and lifetimes of these gases and summaries the literature on Archean sources and concentrations. We recommend the forcings provided here be used both as a~first reference for which gases are likely good greenhouse gases, and as a~standard set of calculations for validation of radiative forcing calculations for the Archean.

\end{abstract}

\introduction

The standard stellar model predicts that the luminosity of a~star increases over its main-sequence lifetime \citep{Gough-1981}. Therefore, the sun is 30\,{\%} brighter now than it was when the solar system formed. Despite a~dimmer sun during the Archean (3.8--2.5~Gyr BP), geologic evidence suggests surface temperatures similar to today \citep{Donn-1965}: This apparent paradox is known as the faint young sun problem (FYSP). To reconcile this, Earth must have had a~lower albedo and/or a~stronger greenhouse effect in the past. In this work we focus on a~stronger greenhouse effect, which is thought to be the primary cause of the warming \citep{Goldblatt-2011a,Wolf-2013}. We focus on the late Archean with a~solar constant of $0.8\,S_0$, resulting in an reduction of $\approx$50\,\unit{W\,m^{-2}} of insolation.

\chem{NH_3} was proposed as a~solution to the FYSP soon after the problem was posed \citep{Sagan-1972}. \chem{NH_3} is a~strong greenhouse gas and concentrations of 10~ppmv could have warmed the Archean surface by 12--15~K \citep{Kuhn-1979}. At the time \chem{NH_3} was proposed as a~solution, it was thought that the early Earth was strongly reducing such that \chem{NH_3} could have built up to significant atmospheric concentrations. However, the Archean atmosphere is now thought to have been only mildly reducing. \chem{NH_3} would likely have photo-dissociated rapidly without UV protection \citep{Kuhn-1979,Kasting-1982}. Furthermore, \chem{NH_3} is highly soluble and would have been susceptible to rain-out. Therefore, sustaining atmospheric concentrations of \chem{NH_3} at which there is significant absorption may not be as easy as originally thought. Considering the destruction of \chem{NH_3} by photolysis, \citet{Kasting-1982} found that concentrations as high as 10~ppbv could plausibly be attained by biotic sources. If \chem{NH_3} were shielded from UV radiation (by a~possible organic haze layer) larger concentrations could be sustained, though concentrations above 1~ppbv seem unlikely \citep{Pavlov-2000}.

 The most obvious resolution to the FYSP would be higher \chem{CO_2} partial pressures. It is believed that the inorganic carbon cycle provides a~strong feedback mechanism which regulates the Earth's temperature over geologic timescales \citep{Walker-1981}. The rate of silicate weathering (a sink of atmospheric \chem{CO_2}) is a~function of surface temperature which depends on the carbon dioxide partial pressure through the greenhouse effect. Therefore, reduced insolation requires higher atmospheric \chem{CO_2} concentrations to regulate the surface temperature and balance the sources (volcanoes) and sinks of atmospheric \chem{CO_2}. However, geological constraints have been proposed which limit atmospheric \chem{CO_2} to levels below those required to keep the early Earth warm \citep{Sheldon-2006,Driese-2011}. \citet{Sheldon-2006} used a~model based on the mass balance of weathering paleosols and find \chem{CO_2} partial pressures between 0.0028-0.026\,\unit{bar} at 2.2~Gyr ago. \citet{Driese-2011} use the same method and find \chem{CO_2} partial pressures between 0.003 and 0.02\,\unit{bar} at 2.69~Gyr ago. However, these constraints are not uniformly accepted \citep{Kasting-2013}.

Other greenhouse gases likely played an important role in the early Earth's energy budget. Most of the focus has been applied to \chem{CH_4} as there are good reasons to expect higher concentrations during the Archean \citep{Zahnle-1986,Kiehl-1987,Pavlov-2000,Haqq-Misra-2008,Wolf-2013}. The Archean atmosphere was nearly anoxic, with very low levels of \chem{O_2}, which would have increased the photochemical lifetime of methane from 10-12 years today to 1000-10,000 years \citep{Kasting-2005}. The concentration of methane in the Archean is not well constrained but \citet{Kasting-2005} suggests that 1-10~ppmv could have been sustained from abiotic sources and up to 1000~ppmv could have been sustained by methanogens. Redox balance models suggest concentrations $\approx$100~ppmv \citep{Goldblatt-2006}. Recent GCM studies have found that reasonably warm climates can be sustained within the bounds of the \chem{CO_2} constraints if the greenhouse is supplemented with elevated \chem{CH_4} concentrations \citep{Wolf-2013,Charnay-2013}. \citet{Wolf-2013} find modern day surface temperature with 0.02\,\unit{bar} of \chem{CO_2} and 1000~ppmv of \chem{CH_4} with 80\,{\%} of present solar luminosity.

Other potential greenhouse gases which have been examined include hydrocarbons \citep{Haqq-Misra-2008}, \chem{N_2O} \citep{Buick-2007,Roberson-2011}, and OCS \citep{Ueno-2009,Hattori-2011}. \chem{C_2H_6} has been suggested to have been radiatively important in the Archean because it can form in significant concentrations from the photolysis of \chem{CH_4} at high partial pressures \citep{Haqq-Misra-2008}. \citet{Haqq-Misra-2008} find that 1~ppmv of \chem{C_2H_6} could increase the surface temp by $\approx$3~K, and 10~ppmv by $\approx$10~K. However, \chem{C_2H_6} is formed along with other organic compounds which form an organic haze. This organic haze is thought to provide a~strong anti-greenhouse effect which limits the utility of \chem{C_2H_6} to warm the climate when produced in this manner.

Elevated OCS concentrations were proposed by \citet{Ueno-2009} to explain the negative $\Delta^{33}$S observed in the Archean sulfate deposits. However, \citet{Hattori-2011} report measurements of ultraviolet OCS absorption cross-sections and find that OCS photolysis does not cause large mass independent fractionation in $\Delta^{33}$S and is therefore not the source of the signatures seen in the geologic record. 

\citet{Buick-2007} proposed that large amounts of \chem{N_2O} could have been produced in the Proterozoic due to bacterial denitrification in copper depleted water, because copper is needed in the enzymatic production of $\mathrm{N_2}$ from \chem{N_2O} (which is the last step of denitrification). \citet{Roberson-2011} find that increasing \chem{N_2O} from 0.3~ppmv to 30~ppmv warms surface temperatures by $\approx8$~K. However, \citet{Roberson-2011} also show that \chem{N_2O} would be rapidly photo-dissociated if \chem{O_2} levels were lower than 0.1~PAL and that \chem{N_2O} was unlikely to have been produced at radiatively important levels at \chem{O_2} levels below this.

Examining the Archean greenhouse involves calculating the radiative effects of greenhouse gases over concentration ranges never before examined. Typically, one time calculations are performed with no standard set of radiative forcings available for comparison. The absence of a~standard set of forcings has led to errors going undetected. For example, the warming exerted by \chem{CH_4} was significantly overestimated by \citet{Pavlov-2000} due to an error in the numbering of spectral intervals \citep{Haqq-Misra-2008} which went undetected for several years. 

 In previous work, greenhouse gas warming has typically been quantified in terms of the equilibrium surface temperature achieved by running a~one-dimensional Radiative-Convective Model (RCM). This metric is sensitive to how climate feedbacks are parametrized in the model and to imposed boundary conditions (e.g., background greenhouse gas concentrations). This makes comparisons between studies and greenhouse gases difficult. It is desirable to document the strengths and relative efficiencies of different greenhouse gases at warming the Archean climate. However, this is near impossible using the literature presently available.

 In this study, we use radiative forcing to quantify changes in the energy budget from changes in greenhouse gas concentrations for a~wide variety of greenhouse gases. We define radiative forcing as the change in the net flux of radiation at the tropopause due to a~change in greenhouse gas concentration with no climate feedbacks. The great utility of radiative forcing is that, to first order, it can be related through a~linear relationship to global mean temperature change at the surface \citep{Hansen-2005}. It therefore provides a~simple and informative metric for understanding perturbations to the energy budget. Furthermore, since radiative forcing is independent of climate response, we get general results which are not affected by uncertainties in the climate response. Radiative forcing has been used extensively to study anthropogenic climate change \citep{IPCC2013}.

Imposed model boundary conditions significantly affect the warming provided by a~greenhouse gas. Boundary conditions that typically vary between studies include: atmospheric pressure, \chem{CO_2} concentrations, and \chem{CH_4} concentrations. The discrepancies in boundary conditions between studies develop from the poorly constrained climatology of the early Earth. In this work, we examine the sensitivity of radiative forcings to variable boundary conditions.

The atmospheric pressure of the Archean is poorly constrained but there are good theoretical arguments to think it was different from today. For one, the atmosphere is 21\,{\%} \chem{O_2} by volume today, whereas there was very little oxygen in the Archean atmosphere. Furthermore, there are strong theoretical arguments that suggest that the atmospheric nitrogen inventory was different: large nitrogen inventories exist in the mantle and continents, which are not primordial and must have ultimately come from the atmosphere \citep{Goldblatt-2009,Johnson-2014}. Constraints on the pressure range have recently been proposed, from raindrop imprints \citep{Som-2012} -- though this has been challenged \citep{Kavanagh-2013}, and from noble gas systematics \citep[0.7-1.1 bar,][]{Bernard-2013}.

Atmospheric pressure affects the energy budget in two ways. (1) Increasing pressure increases the moist adiabatic lapse rate. The moist adiabatic lapse rate is a~function of the saturation mixing ratio of water vapour. The saturation vapour pressure is independent of pressure. Increasing pressure means there is more dry air to absorb the latent heat released by condensation, making the moist adiabatic lapse rate larger (closer to the dry adiabatic lapse rate). (2) As the pressure increases, collisions between molecules become more frequent. This results in a~broadening of the absorption lines over a~larger frequency range. This phenomena is called pressure broadening and generally causes more absorption \citep{Goody-1995}. 

Changes to the concentrations of \chem{CO_2} and \chem{CH_4} will affect the strength of other greenhouse gases. When multiple gases absorb radiation at the same frequencies, the total absorption is less than the sum of the absorption that each gas contributes in isolation. This difference is known as overlap. It occurs because the absorption is distributed between the gases, so in effect there is less radiation available for each gas to absorb.

In this paper, we present calculations of radiative forcings for \chem{CO_2}, \chem{CH_4} and 26 other gases contained in the HIgh-resolution TRANsmission (HITRAN) molecular database for atmospheres with 0.5\,\unit{bar}, 1\,\unit{bar}, and 2\,\unit{bar} of $\mathrm{N_2}$. We aim to provide a~complete set of radiative forcing and overlap calculations which can be used as a~standard for comparisons. We provide \chem{CO_2} and \chem{CH_4} radiative forcings over large ranges in concentration, and compare our results with calculations in the literature. For the other 26 HITRAN gases, the HITRAN absorption data is compared with measured cross-sections and discrepancies are documented. Radiative forcings are calculated over a~concentrations range of 10~ppbv to 10~ppmv. The sensitivites of the radiative forcings to atmospheric pressure and overlapping absorption with other gases are examined, and our results are compared with results from the literature.

This paper is organized as follows. In section 2, we describe our general methods, evaluation of the spectral data and the atmospheric profile we use. In section 3, we examine the radiative forcings due to \chem{CO_2} and \chem{CH_4} and examine how our results compare with previous calculations. In section 4, we provide radiative forcings for 26 other gases from the HITRAN database and examine the sensitivity of these results to atmospheric parameters.

\section{Methods}
\subsection{Overview}

We calculate absorption cross-sections from HITRAN line parameters and compare our results with measured cross-sections. We develop a~single-column atmospheric profile based on constraints of the Archean atmosphere. With this profile, we perform radiative forcing calculations for \chem{CO_2}, \chem{CH_4} and 26 other HITRAN gases. Gas amounts are given in abundances, $a$, relative to the modern atmosphere (1\,\unit{bar}, molecular weight of 28.97~g/moles, total moles ($n_0$) of $\approx1.8{\times}10^{20}$). Thus $a=n_{gas}/n_0$. As an example, an abundance of 1 for \chem{CO_2} contains the same number of moles as the modern atmosphere but would give a~surface pressure larger than 1\,\unit{bar} because of the higher molecular weight. For our experiments we add gas abundances to background $\mathrm{N_2}$ partial pressure, increasing the atmospheric pressure.

\subsection{Spectra}

Line parameters are taken from the HITRAN 2012 database \citep{Rothman-2013}. We use the LBLABC code, written by David Crisp, to calculate cross-sections from the line date. Line parameters have a~significant advantage over measured absorption cross-sections, in that, absorption can be calculated explicitly as a~function of temperature and pressure. The strength of absorption lines is a~function of temperature and shape is a~function of pressure. Neglecting these dependencies can result in significant errors in radiative transfer calculations.

There are, however, some limitations to using HITRAN data. \citet{Rothman-2009} explains that the number of transitions included in the database is limited by: (1) a~reasonable minimum cutoff in absorption intensity (based on the sensitivity of instruments that observe absorption over extreme terrestrial atmospheric path lengths), (2) lack of sufficient experimental data, or (3) lack of calculated transitions. The molecules for which data are included in the line-by-line portion of HITRAN are mostly composed of small numbers of atoms and have low molecular weights. Large polyatomic molecules have many normal modes of vibration and have fundamentals at very low wavenumbers \citep{Rothman-2009}. This makes it difficult to experimentally isolate individual lines of large molecules, so that a~complete set of line parameters for these molecules is impossible to obtain.

    \begin{figure*}
\includegraphics[height=100mm]{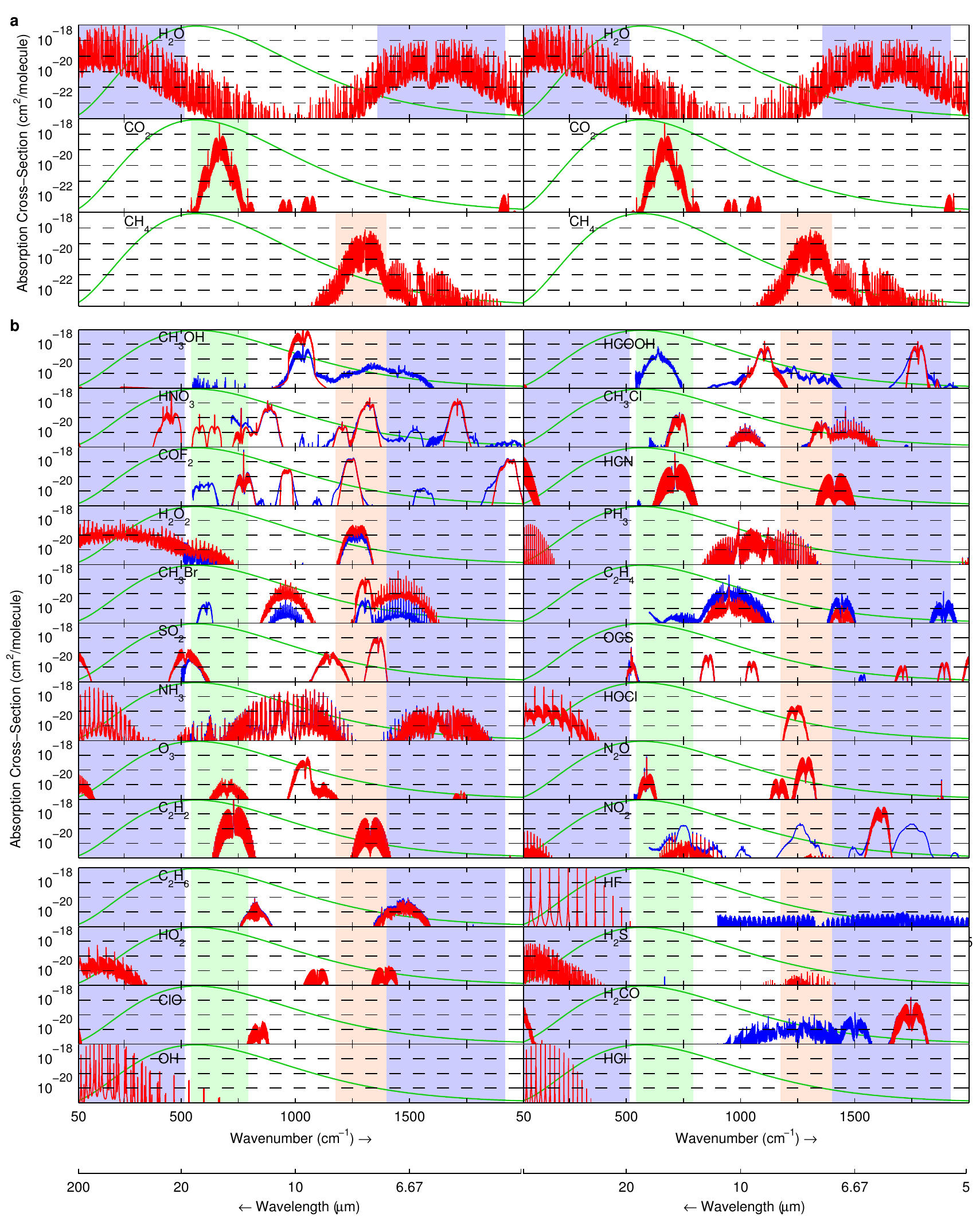}
 \caption{\textit{Absorption cross-sections.} Absorption cross-sections of (a) \chem{H_2O}, \chem{CO_2}, and \chem{CH_4} and (b) potential early Earth trace gases. Cross-sections are calculated from HITRAN line data (red) and measured from the PNNL database (blue) at 1013 hPa and 278 K. The gases are ordered from strongest to weakest based on the analysis in section \ref{section:TraceGases} (Fig.~\ref{fig:absorption}) in columns from top left to bottom right. Colored shaded areas show wavenumbers at which absorption is strongest for \chem{H_2O} (blue), \chem{CO_2} (green), and \chem{CH_4} (red). The green curve shows the shape of the blackbody emissions from a~289~K blackbody.}
 \label{fig:xsec1}
 \end{figure*}

Computed cross-sections are compared to measured cross-sections from the Pacific Northwest National Laboratory (PNNL) database \citep{Sharpe-2004} for the strongest HITRAN gases (Fig.~\ref{fig:xsec1}). Where differences exist, it is not straight forward to say which is in error (for example, potential problems with measurements include contamination of samples). 
Hence we simply note any discrepancy and do our best to note the consequences of these. The largest abundance of the trace gases examined in this work is $10^{-5}$, at this abundance only absorption cross-section greater than $\approx 5 \times 10^{-21} \; \mathrm{cm^2}$ absorb strongly over the depth of the atmosphere.

The similarities and differences between the cross-sections for each
gas are:
\begin{itemize}
\item \chem{CH_3OH}: The HITRAN line data covers the range of 975--1075~$\mathrm{cm^{-1}}$. In that range the HITRAN cross-sections are an order of magnitude larger than the PNNL cross-sections. Therefore, the PNNL data suggests the abundances should be an order of magnitude larger to obtain the same forcings as the HITRAN data. PNNL cross-sections indicate that there is missing HITRAN line data over the range 1075--1575~$\mathrm{cm^{-1}}$ with peaks of $\approx5\times10^{-20}\,\mathrm{cm^2}$, which would be optically thick for abundances $\geq10^{-6}$. There is also missing HITRAN data at 550--750~$\mathrm{cm^{-1}}$ with peaks of $\approx5\times10^{-21}\,\mathrm{cm^2}$, which would be optically thick for abundances $\geq10^{-5}$.
\item \chem{HNO_3}: The HITRAN data covers the range of 550--950~$\mathrm{cm^{-1}}$, 1150--1400~$\mathrm{cm^{-1}}$, and 1650--1750~$\mathrm{cm^{-1}}$. Over this range, HITRAN and PNNL data agree well except between 725-825~$\mathrm{cm^{-1}}$ where the PNNL cross-sections are larger (relevant for abundances of $\geq5\times10^{-7}$). PNNL cross-sections indicate that there is significant missing HITRAN line data in the ranges 1000--1150~$\mathrm{cm^{-1}}$, 1400--1650~$\mathrm{cm^{-1}}$, and 1750--2000~$\mathrm{cm^{-1}}$, which would be optically thick for abundances $\geq5\times10^{-6}$.
\item \chem{COF_2}: The HITRAN cross-sections cover the range of 725--825~$\mathrm{cm^{-1}}$, 950--1000~$\mathrm{cm^{-1}}$, 1175--1300~$\mathrm{cm^{-1}}$, and 1850--2000~$\mathrm{cm^{-1}}$. The PNNL and HITRAN cross-sections agree over this range. HITRAN is missing bands around 650 and 1600~$\mathrm{cm^{-1}}$ with peaks of $\approx10^{-20}\,\mathrm{cm^2}$, which would be optically thick for abundances $\geq5\times10^{-6}$. Additionally, wings of 950--1000~$\mathrm{cm^{-1}}$, 1175--1300~$\mathrm{cm^{-1}}$, and 1850--2000~$\mathrm{cm^{-1}}$ bands appear missing in HITRAN, relevant at similar abundances.
\item \chem{H_2O_2}: Above 500~$\mathrm{cm^{-1}}$, the HITRAN and PNNL cross-sections cover the same wavenumber range. Over this range, HITRAN cross-sections are about twice the value of the PNNL cross-sections. Therefore, the PNNL data suggests the abundances should be about twice those of the HITRAN data to obtain the same forcings.
\item \chem{CH_3Br}: HITRAN cross-sections are over an order of magnitude greater than the PNNL cross-sections. Therefore, the PNNL data suggests the abundances should be $\approx13$ times those of the HITRAN data to obtain the same forcings. The PNNL cross-sections indicate missing HITRAN line data over the range of 575--650~$\mathrm{cm^{-1}}$ with peaks of $\approx10^{-20}\,\mathrm{cm^2}$, which would be optically thick for abundances $\geq5\times10^{-6}$.
\item \chem{SO_2}: The HITRAN and PNNL cross-sections agree well except between 550--550~$\mathrm{cm^{-1}}$ where HITRAN cross-sections are larger with peaks of $\approx5\times10^{-20}\,\mathrm{cm^2}$, which would be optically thick for abundances $\geq10^{-7}$.
\item \chem{NH_3}: the HITRAN and PNNL cross-sections agree well.
\item \chem{O_3}: there is no PNNL data for this gas.
\item \chem{C_2H_2}: the HITRAN and PNNL cross-sections agree well.
\item \chem{HCOOH}: The HITRAN data between 1000--1200 and 1725--1875~$\mathrm{cm^{-1}}$ agrees with the PNNL data. The PNNL cross-sections indicate missing line data over the range 550--1000~$\mathrm{cm^{-1}}$ with peaks of $\approx5\times10^{-19}\,\mathrm{cm^2}$, which would be optically thick for abundances $\geq10^{-8}$., 1200--1725~$\mathrm{cm^{-1}}$, and 1875--2000~$\mathrm{cm^{-1}}$ with peaks of $\approx5\times10^{-20}\,\mathrm{cm^2}$, which would be optically thick for abundances $\geq10^{-7}$.
\item \chem{CH_3Cl}: The HITRAN and PNNL cross-sections agree well. PNNL cross-sections indicate missing line data around 600~$\mathrm{cm^{-1}}$.
\item \chem{HCN}: the HITRAN and PNNL cross-sections agree well.
\item \chem{PH_3}: the HITRAN and PNNL cross-sections agree well.
\item \chem{C_2H_4}: The HITRAN data is about an order of magnitude less than PNNL. Therefore, the PNNL data suggests the abundances should be an order of magnitude less to obtain the same forcings as the HITRAN data.
\item \chem{OCS}: the HITRAN and PNNL cross-sections agree well.
\item \chem{HOCl}: there is no PNNL data for this gas.
\item \chem{N_2O}: the HITRAN and PNNL cross-sections agree well.
\item \chem{NO_2}: The HITRAN and PNNL cross-sections agree well in the range 1550--1650~$\mathrm{cm^{-1}}$. The PNNL cross-sections are up to an order of magnitude larger than HITRAN for cross-sections in the range 650--850~$\mathrm{cm^{-1}}$ and around 1400~$\mathrm{cm^{-1}}$. PNNL cross-sections indicate missing line data over the ranges 850-1100~$\mathrm{cm^{-1}}$ and 1650--2000~$\mathrm{cm^{-1}}$ with peaks of $\approx10^{-19}\,\mathrm{cm^2}$, which would be optically thick for abundances $\geq5\times10^{-7}$.
\item \chem{C_2H_6}: the HITRAN and PNNL cross-sections agree well.
\item \chem{HO_2}: there is no PNNL data for this gas.
\item \chem{ClO}: there is no PNNL data for this gas.
\item \chem{OH}: there is no PNNL data for this gas.
\item $\mathrm{HF}$: The HITRAN and PNNL data do not overlap. The HITRAN data is available below 500~$\mathrm{cm^{-1}}$ and PNNL data is available above $\approx$900~$\mathrm{cm^{-1}}$.
\item $\mathrm{H_2S}$: The HITRAN and PNNL cross-sections agree well in the range 1100--1400~$\mathrm{cm^{-1}}$. There is no PNNL data for the absorption feature at wavenumbers less that 400~$\mathrm{cm^{-1}}$.
\item $\mathrm{H_2CO}$: The HITRAN and PNNL cross-sections agree well for the absorption band in the range 1600--1850~$\mathrm{cm^{-1}}$. The HITRAN data is missing the absorption band over the wavenumber range 1000--1550~$\mathrm{cm^{-1}}$ with peaks of $\approx10^{-20}\,\mathrm{cm^2}$, which would be optically thick for abundances $\geq5\times10^{-6}$.
\item \chem{HCl}: there are no optically thick absorption features
  over the wavenumbers where PNNL data exists.
\end{itemize}

The spectral data described above only covers the thermal
spectrum. HITRAN line parameters are not available for the solar
spectrum (other than \chem{CO_2}, \chem{CH_4}, \chem{H_2O}, and
\chem{O_3}). We are unaware of any absorption data for these gases in
the solar spectrum. If these gases are strong absorbers in the solar
spectrum (e.g.~\chem{O_3}) the radiative forcing calculations could
be significantly affected. Very strong heating in the stratosphere
would cause dramatic differences in the stratospheric structure which
would significantly affect the radiative forcing.

\subsection{Atmospheric profile}

We perform our calculations for a~single-column atmosphere. Performing
radiative forcing calculations for a~single profile rather than
multiple profiles representing the meridional variation in the Earth's
climatology introduces only small errors
\citep{Myhre-1997,Freckleton-1998,Byrne-2014}.

The tropospheric temperature structure is dictated largely by convection. We approximate the tropospheric temperature structure with the pseudo-adiabatic lapse rate. The lapse rate is dependent on both pressure and temperature. There is a~large range of uncertainty in the surface temperatures of the Archean, we take the surface temperature to be the Global and Annual Mean (GAM) temperature on the modern Earth (289~K). We chose this temperature for two reasons. (1) It makes comparisons with the modern Earth straight forward. (2) Glaciations appear rare in the Archean \citep{Young-1991}, thus, it is expected that surface temperatures were likely as warm as today for much of the Archean. Therefore, modern day surface temperatures are a~reasonable assumption for our profile.

We calculate three atmospheric profiles for $\mathrm{N_2}$ inventories of 0.5\,\unit{bar}, 1\,\unit{bar}, and 2\,\unit{bar}. Atmospheric pressure varies with the addition of \chem{CO_2} and \chem{CH_4}. We use the GAM relative humidity from Modern Era Retrospective-analysis for Research and Applications reanalysis data products \citep{Rienecker-2011} over the period 1979 to 2011.

   \begin{figure*}
\includegraphics[width=90mm]{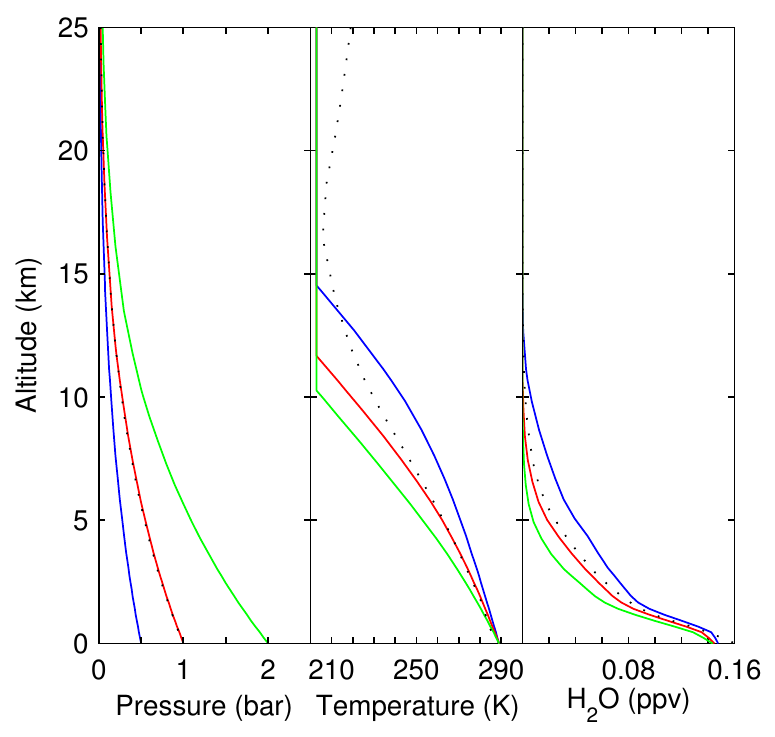}
 \caption{\textit{Atmospheric Profiles.} Pressure, temperature and water vapor structure of atmospheres with 0.5\,\unit{bar} (blue), 1\,\unit{bar} (red), and 2\,\unit{bar} (green) of $\mathrm{N_2}$. The modern atmosphere is also shown (dotted). The water vapor abundances are scaled to an atmosphere with 1 bar of $\mathrm{N_2}$.}
 \label{fig:profiles}
 \end{figure*}
 
In contrast to the troposphere, the stratosphere (taken to be from the tropopause to the top of the atmosphere) is near radiative equilibrium. The stratospheric temperature structure is therefore sensitive to the abundances of radiatively active gases. %If the stratosphere is optically thin and
For a~grey gas, an optically thin stratosphere heated by upwelling radiation will be isothermal at the atmospheric skin temperature \citep[$T=\left(I(1-\alpha)/8 \sigma \right)^{1/4} \approx 203\, \mathrm{K},$][]{Pierrehumbert-2010}. We take this to be the case in our calculations. In reality, non-grey gases can give a~warmer or cooler stratosphere depending on the spectral positioning of the absorption lines. Furthermore, the stratosphere would not have been optically thin, as \chem{CO_2} (and possibly other gases) were likely optically thick for some wavelengths, which would have cooled the stratosphere. Other gases, such as \chem{CH_4}, may have significantly warmed the stratosphere by absorbing solar radiation. However, the abundances of these gases are poorly constrained. Since there is no convincing reason to choose any particular profile, we keep the stratosphere at the skin temperature for simplicity. The atmospheric profiles are shown in figure \ref{fig:profiles}. We take the tropopause as the atmospheric level at which the pseudoadiabatic lapse rate reaches the skin temperature. Sensitivity tests were performed to examine the sensitivity of radiative forcing to the temperature and water vapour structure. We find that differences in radiative forcing are generally small ($\le 10\,{\%}$, appendix \ref{section:appendix}).

In this study, we explicitly include clouds in our radiative transfer calculations. Following \citet{Kasting-1984}, many RCMs used to study the Archean climate have omitted clouds, and adjusted the surface albedo such that the modern surface temperatures can be achieved with the current atmospheric composition and insolation. \citet{Goldblatt-2011a} showed that neglecting the effects of clouds on longwave radiation can lead to significant over-estimates of radiative forcings, as clouds absorb longwave radiation strongly and with weak spectral dependence. Clouds act as a~new surface of emission to the top of the atmosphere and, therefore, the impact on the energy budget of molecular absorption between clouds and the surface is greatly reduced. We take our cloud climatology as cloud fractions and optical depths from International Satellite Cloud Climatology Project D2 data set, averaging from January 1990 to December 1992. This period is used by \citet{Rossow-2005} and was chosen so that we could compare cloud fractions. We assume random overlap and average by area to estimate cloud fractions. The clouds were placed at 226~K for high clouds, 267~K for middle clouds and 280~K for low clouds, this corresponds to the average temperature levels of clouds on the modern Earth. Cloud properties are taken from \citet{Byrne-2014}. The cloud climatology of the Archean atmosphere is highly uncertain. Recent GCM studies have found that there may have been less cloud cover due to less surface heating from reduced insolation \citep{Charnay-2013,Wolf-2013}. Other studies have suggested other mechanisms which could have caused significant changes in cloud cover during the Archean \citep{Rondanelli-2010,Rosing-2010,Shaviv-2003} although theoretical problems have been found with all of these studies \citep{Goldblatt-2011a,Goldblatt-2011-Nat}. Nevertheless, given the large uncertainties in the cloud climatology in the Archean the most straight forward assumption is to assume modern climatology, even though there were likely differences in the cloud climatology. Furthermore, the goal of this study is to examine greenhouse forcings and not cloud forcings. Therefore, we want to capture the longwave effects of clouds to a~first order degree. Differences in cloud climatology have only secondary effects on the results given here.

Atmospheric profiles are provided as supplementary material.

\subsection{Radiative forcing calculations}

We use the Spectral Mapping for Atmospheric Radiative Transfer (SMART) code, written by David Crisp \citep{Meadows-1996}, for our radiative transfer calculations. This code works at line-by-line resolution but uses a~spectral mapping algorithm to treat different wavenumber regions with similar optical properties together, giving significant savings in computational cost. We evaluate the radiative transfer in the range 50--100,000~$\mathrm{cm^{-1}}$ (0.1--200~$\mathrm{\mu m}$) as a~combined solar and thermal calculation.

Radiative forcing is calculated by performing radiative transfer calculations on atmospheric profiles with perturbed and unperturbed greenhouse gas abundances and taking the difference in net flux of radiation at the tropopause. We assume the gases examined here are well-mixed.

\section{Results and discussion}
\subsection{\chem{CO_2}}

     \begin{figure*}
\includegraphics[width=90mm]{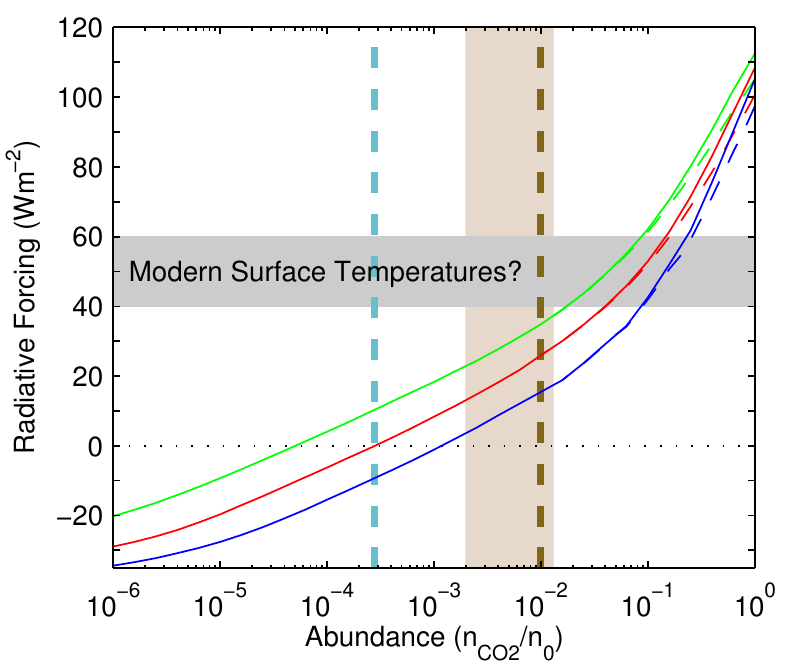}
 \caption{ \textit{\chem{CO_2} Radiative Forcings.} Radiative forcing as a~function of \chem{CO_2} abundance, relative to pre-industrial \chem{CO_2} (1\,\unit{bar} $\mathrm{N_2}$). Colors are for atmospheres with 0.5\,\unit{bar}, 1\,\unit{bar}, and 2\,\unit{bar} of $\mathrm{N_2}$. Solid lines are calculated with CIA and dashed lines are calculated without CIA. The shaded region shows the range of \chem{CO_2} for the early Earth \citep[0.003-0.02\,\unit{bar},][]{Driese-2011}. The vertical dashed blue and brown lines give the pre-industrial and early Earth best guess ($10^{-2}$) abundances of \chem{CO_2}.}
 \label{fig:CO2forcing}
 \end{figure*}

We calculate \chem{CO_2} radiative forcings up to an abundance of 1 (Fig.~\ref{fig:CO2forcing}). At $10^{-2}$, consistent with paleosol constraints, the radiative forcings are 35\,\unit{W\,m^{-2}} (2\,\unit{bar} $\mathrm{N_2}$), 26\,\unit{W\,m^{-2}} (1\,\unit{bar} $\mathrm{N_2}$), and 15\,\unit{W\,m^{-2}} (0.5\,\unit{bar} $\mathrm{N_2}$), which is considerably short of the forcing required to solve the FYSP, consistent with previous work. The \chem{CO_2} forcings given here account for changes in the atmospheric structure due to changes in the $\mathrm{N_2}$ inventory and thus are non-zero at pre-industrial \chem{CO_2} for 0.5\,\unit{bar} and 2\,\unit{bar} of $\mathrm{N_2}$. This results in forcings of about 10\,\unit{W\,m^{-2}} (2\,\unit{bar}) and -9\,\unit{W\,m^{-2}} (0.5\,\unit{bar}) at pre-industrial \chem{CO_2} abundances \citep[see][for a~detailed physical description]{Goldblatt-2009}.

 At very high \chem{CO_2} abundances ($>0.1$), \chem{CO_2} becomes a~significant fraction of the atmosphere. This complicates radiative forcing calculations by (1) changing the atmospheric structure, (2) shortwave absorption/scattering, and  (3) uncertainties in the parametrization of continuum absorption. These need careful consideration in studies of very high atmospheric \chem{CO_2}, so we describe these factors in detail:

  \begin{figure*}
\includegraphics[height=100mm]{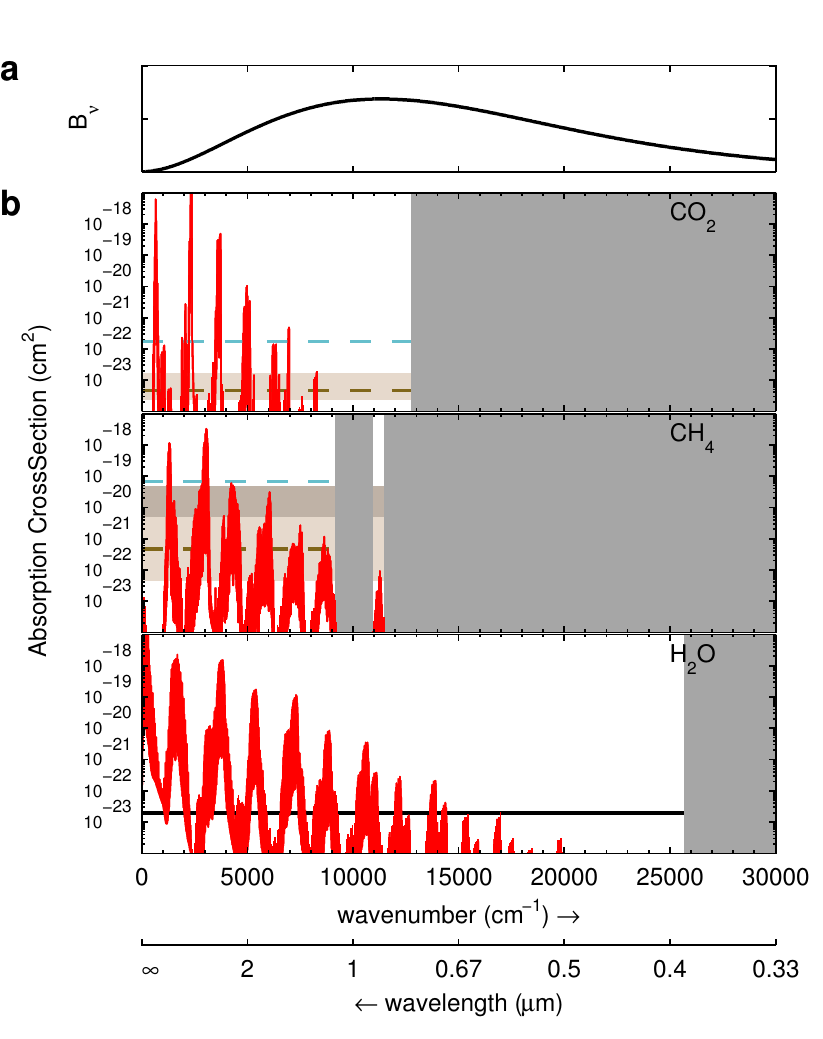}
 \caption{\textit{Solar absorption cross-sections}. (a) Emission spectrum for an object of 5777~K (Effective emitting temperature of modern Sun). (b) Absorption cross-sections of \chem{CO_2}, \chem{CH_4} and \chem{H_2O} calculated from line data. Grey shading shows were there is no HITRAN line data. Shaded and dashed lines show absorption cross-sections of unity optical depth for abundances given in figures \ref{fig:CO2forcing} and \ref{fig:CH4forcing}. Solid black line shows the absorption cross-sections of unity optical depth for \chem{H_2O}.}
 \label{fig:solarspectra}
 \end{figure*}
 
\begin{enumerate}
\item Large increases in \chem{CO_2} increase the atmospheric pressure, and therefore, also increase the atmospheric lapse rate. This results in a~cooling of the stratosphere and reduction to the emission temperature. Increased atmospheric pressure also results in the broadening of absorption lines for all of the radiatively active gases. Emissions from colder, higher pressure, layers increases radiative forcing.
\item Shortwave radiation is also affected by very high \chem{CO_2} abundances. The shortwave forcing is -2\,\unit{W\,m^{-2}} at 0.01, -4\,\unit{W\,m^{-2}} at 0.1 and -18\,\unit{W\,m^{-2}} at 1. There are two separate reasons for this. The smaller affect is absorption of shortwave radiation by \chem{CO_2} which primarily affects wavenumbers less than $\approx10,000 \, \mathrm{cm^{-1}}$ (Fig.~\ref{fig:solarspectra}). The most important effect ($\mathrm{CO_2}>$ 0.1) is increased Rayleigh scattering due to the increase in the size of the atmosphere. This primarily affects wavenumbers larger than $\approx10,000 \, \mathrm{cm^{-1}}$ and is the primary reason for the large difference in insolations through the tropopause for abundances between 0.1 and 1 of \chem{CO_2}.
\item There is significant uncertainty in the \chem{CO_2} spectra at very high abundances. This is primarily due to absorption that varies smoothly with wavenumber that cannot be accounted for by nearby absorption lines. This absorption is termed continuum absorption and is caused by the far wings of strong lines and collision induced absorption (CIA) \citep{Halevy-2009}. \citet{Halevy-2009} show that different parametrizations of line and continuum absorption in different radiative transfer models can lead to large differences in outgoing longwave radiation at high \chem{CO_2} abundances. SMART treats the continuum by using a~$\chi$-factor to reduce the opacity of the Voight line shape out to $1000 \; \mathrm{cm^{-1}}$ from the line center to match the background absorption. We add to this CIA absorption which has been updated with recent results of \citet{Wordsworth-2010}. We believe that our radiative transfer runs are as accurate as possible given the poor understanding of continuum absorption.
\end{enumerate}

It is worthwhile comparing our calculated radiative forcings with previous results. In most studies, the greenhouse warming from a~perturbation in greenhouse gas abundance is quantified as a~change of the GAM surface temperature. We convert our radiative forcings to surface temperatures for comparison. This is achieved using climate sensitivity. Assuming the climate sensitivity to be in the range 1.5--4.5\,\unit{W\,m^{-2}} \citep[medium confidence range,][]{IPCC2013} for a~doubling of atmospheric \chem{CO_2} and the radiative forcing for a~doubling of \chem{CO_2} to be $\mathrm{3.7\;Wm^{-2}}$, we find a~range of climate sensitivity parameters of 0.4--1.2~$\mathrm{K/Wm^{-2}}$ with a~best guess of 0.8~$\mathrm{K/Wm^{-2}}$. We take the \chem{CO_2} abundance which gives energy balance at the tropopause (0.20\,\unit{bar}, abundance of 0.13) to be the abundance that gives a~surface temperature of 289~K.

     \begin{figure*}
\includegraphics[width=90mm]{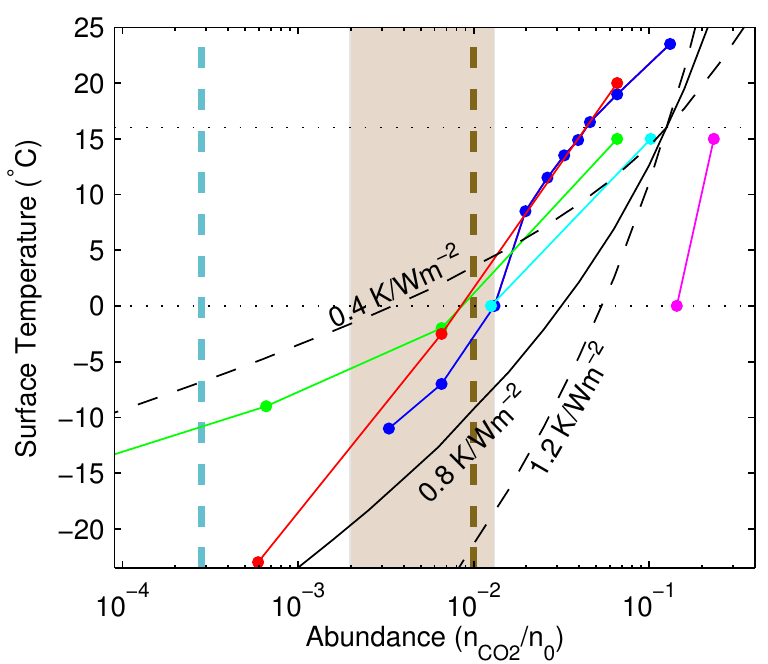}
 \caption{\textit{Surface temperature as a~function of \chem{CO_2} abundance for $\mathit{0.8 \, S_0}$}. Temperatures are calculated from radiative forcings assuming climate sensitivity parameters of 0.4~$\mathrm{K/Wm^{-2}}$ (dashed black), 0.8~$\mathrm{K/Wm^{-2}}$ (solid black) and 1.2~$\mathrm{K/Wm^{-2}}$ (dashed black) and a~surface temperature of 289~K at 0.20\,\unit{bar} of \chem{CO_2} (when our model is in energy balance). The results of \citet{Wolf-2013} (blue), \citet{Haqq-Misra-2008} (green), \citet{Charnay-2013} (red), \citet{vonParis-2008} (cyan), and \citet{Kienert-2012} (magenta) are also shown.}
 \label{fig:CO2Temp}
 \end{figure*}
 
The calculated temperature curves are plotted with the results of previous studies (Fig.~\ref{fig:CO2Temp}). For all of the studies, surface temperatures were calculated for 0.8$\mathrm{S_0}$. However there were differences in the atmospheric pressure: \citet{vonParis-2008} and \citet{Kienert-2012} have 0.77\,\unit{bar} and 0.8\,\unit{bar} of $\mathrm{N_2}$ respectively, while \citet{Haqq-Misra-2008}, \citet{Wolf-2013} and \citet{Charnay-2013} hold the surface pressure at 1\,\unit{bar}, and remove $\mathrm{N_2}$ to add \chem{CO_2}.

Model climate sensitivities can be grouped by the type of climate model used. Simple 1-D RCMs \citep{Haqq-Misra-2008,vonParis-2008} have the lowest climate sensitivities (1-4~K). The 3-D models had higher climate sensitivities, but the sensitivities were also more variable between models. \citet{Kienert-2012} use a~model with a~fully dynamic ocean but a~statistical dynamical atmosphere. The sea-ice albedo feedback makes the climate highly sensitive to \chem{CO_2} abundance and has the largest climate sensitivity ($\approx$18.5~K). \citet{Charnay-2013} and \citet{Wolf-2013} use models with fully dynamic atmosphere but with simpler oceans. They generally have climate sensitivities between 2.5-4.5~K but \citet{Wolf-2013} find higher climate sensitivities (7-11~K) for \chem{CO_2} concentrations of 10,000--30,000~ppmv due to changes in surface albedo (sea ice extent). The climate sensitivities are larger for the 3-D models compared to the RCMs primarily because of the ice-albedo feedback. Variations in climate sensitivity parameters mask variations in radiative forcings.

The amount of \chem{CO_2} required to reach modern day surface temperatures is variable between models. \citet{Charnay-2013} and \citet{Wolf-2013} require the least \chem{CO_2} to sustain modern surface temperatures (0.06--0.07\,\unit{bar}, abundance of 0.04--0.46), primarily because there are less clouds (low and high), the net effect of which is a~decrease in albedo. The cloud feedback in these models works as follows: the reduced insolation results in less surface heating, which results in less evaporation and less cloud formation. The RCM studies require \chem{CO_2} abundance very close to our results (0.1-0.2\,\unit{bar}, abundance of 0.066--0.13), especially considering differences in atmospheric pressure. \citet{Kienert-2012} requires very high \chem{CO_2} abundances ($\approx$0.4\,\unit{bar}, abundance of 0.26) to prevent runaway glaciation because of the high sensitivity of the ice-albedo feedback in this model.

\subsection{\chem{CH_4}}

     \begin{figure*}
\includegraphics[width=90mm]{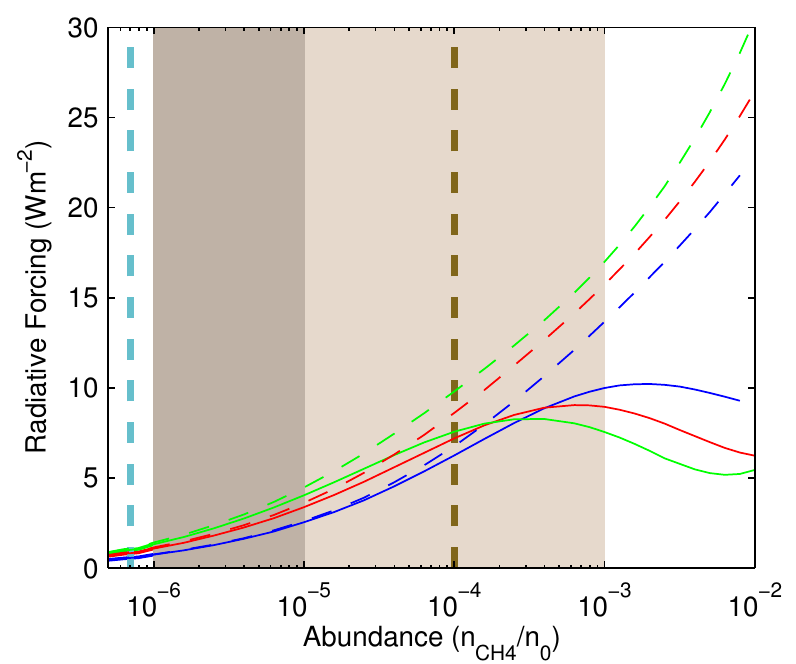}
 \caption{ \textbf{Radiative Forcing for \chem{CH_4}.} Radiative forcing as a~function of \chem{CH_4} for atmospheres with 0.5\,\unit{bar} (blue), 1\,\unit{bar} (red) and 2\,\unit{bar} (green) of $\mathrm{N_2}$. Dashed curves show the longwave forcing. Shaded region shows the range of \chem{CH_4} for the early Earth that could be sustained by abiotic (dark) and biotic (light) sources \citep{Kasting-2005}. The vertical dashed blue and brown lines give the pre-industrial and early Earth best guess \citep[100~ppmv,][]{Goldblatt-2006} abundances of \chem{CH_4}.}
 \label{fig:CH4forcing}
 \end{figure*}

      \begin{figure*}
\includegraphics[width=90mm]{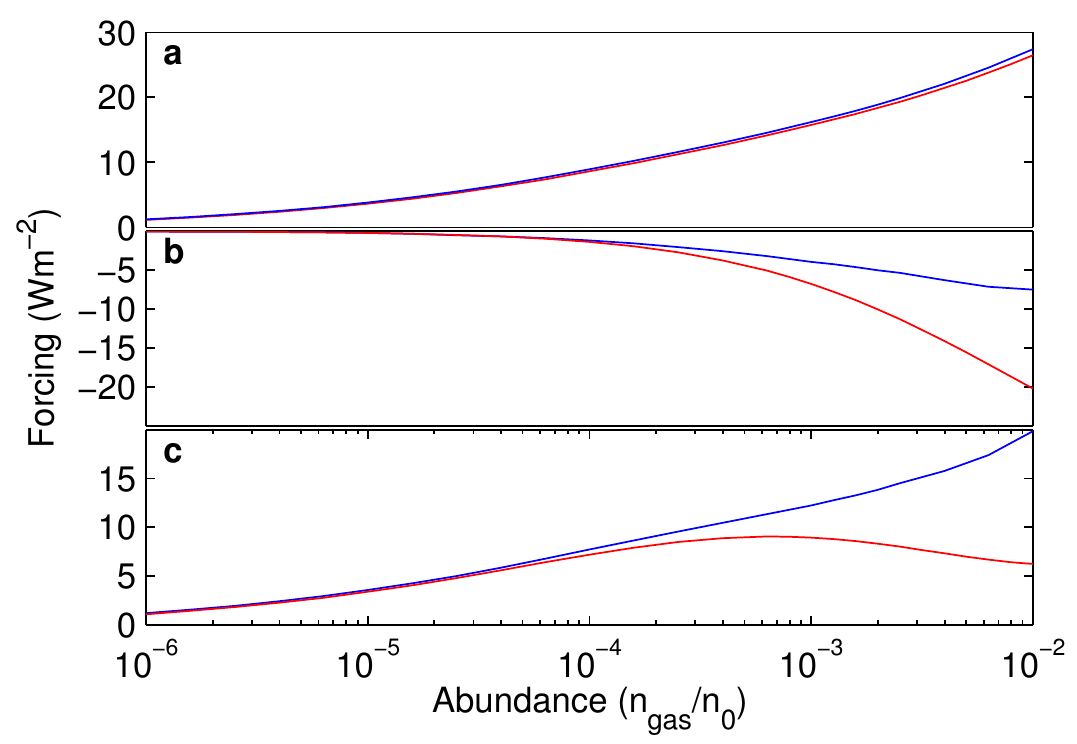}
       \caption{\chem{CH_4} \textit{forcings using HITRAN 2000 and
           2012 spectral data.} \textbf{(a)} Longwave, \textbf{(b)}
         Shortwave, and \textbf{(c)} combined longwave and shortwave
         radiative forcings using HITRAN 2000 (blue) and 2012 (red)
         spectral data.}
 \label{fig:forcing_2kand2012}
 \end{figure*}
 
    \begin{figure*}
\includegraphics[width=120mm]{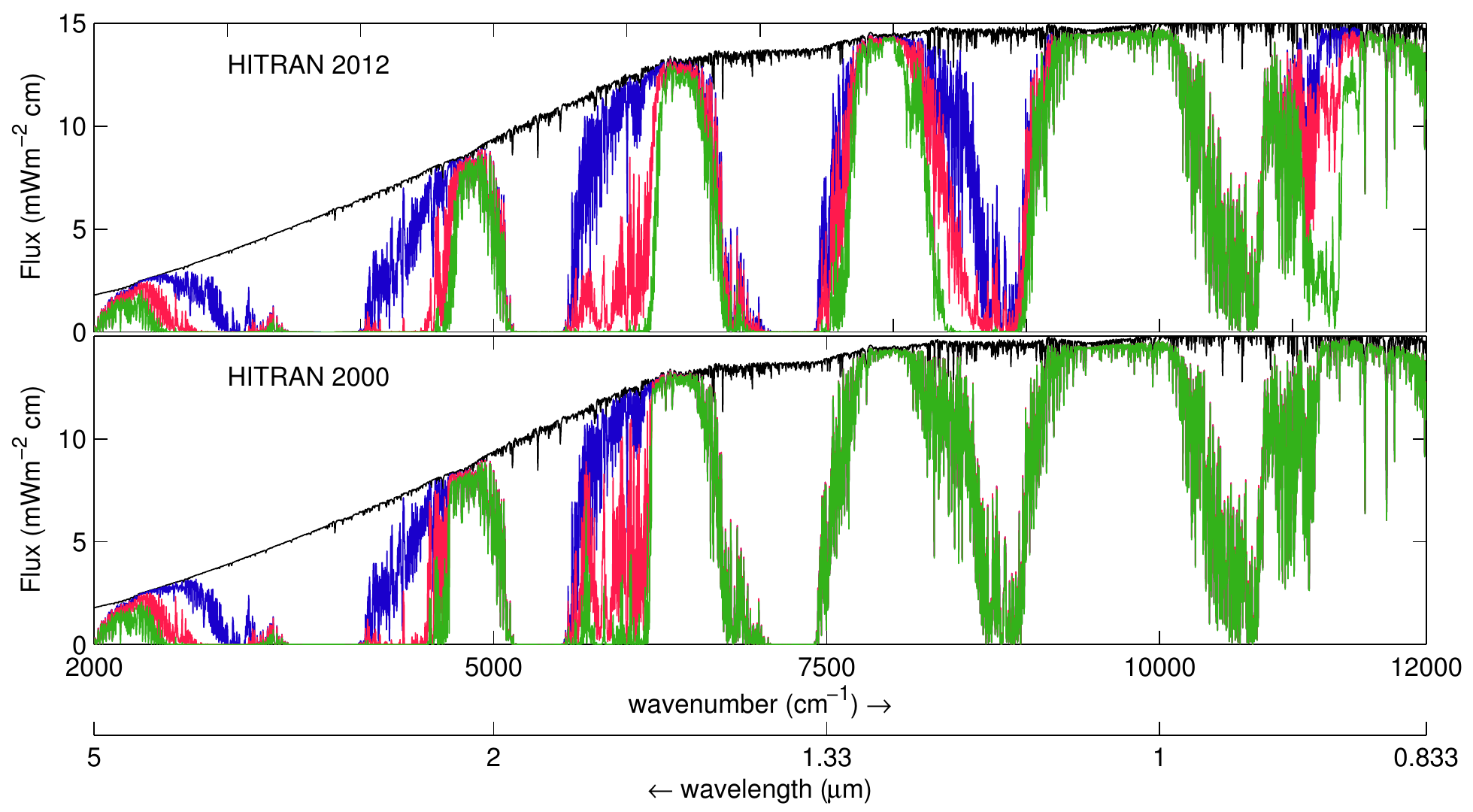}
     \caption{\textit{Downward shortwave flux.} Insolation at the top of the atmosphere (black) and surface for \chem{CH_4} abundances $10^{-6}$ (blue), $10^{-4}$ (red), and $10^{-2}$ (green) using HITRAN 2012 (top) and HITRAN 2000 (bottom) line data.}
 \label{fig:solarabsorption}
 \end{figure*}
 
We calculate \chem{CH_4} radiative forcings up to $10^{-2}$ (Fig.~\ref{fig:CH4forcing}). At abundances greater than $10^{-4}$ we find considerable shortwave absorption. For an atmosphere with 1\,\unit{bar} of $\mathrm{N_2}$, the shortwave radiative forcing is $\approx 1.4 \,\mathrm{Wm^{-2}}$ at $10^{-4}$, $\approx 6.7 \,\mathrm{Wm^{-2}}$ at $10^{-3}$, and $\approx 20 \,\mathrm{Wm^{-2}}$ at $10^{-2}$. This absorption occurs primarily at wavenumbers less than 11,502~$\mathrm{cm^{-1}}$ where HITRAN data is available (Fig.~\ref{fig:solarspectra}). We find much more shortwave absorption here than in studies from Jim Kasting's group \citep{Pavlov-2000,Haqq-Misra-2008} which also parametrize solar absorption. The reason for this discrepancy is likely due to improvements in the spectroscopic data. We repeated our radiative forcing calculations using HITRAN 2000 line parameters and found only minor differences in the longwave absorption but much less absorption at solar wavelengths (Fig.~\ref{fig:forcing_2kand2012}). Figure \ref{fig:solarabsorption} shows the solar absorption by \chem{CH_4} using spectra from the 2000 and 2012 editions of HITRAN. There is a~significant increase in shortwave absorption between 5500--9000~$\mathrm{cm^{-1}}$ and around 11,000~$\mathrm{cm^{-1}}$.

 Very strong shortwave absorption would have a~significant effect on the temperature structure of the stratosphere. Strong absorption would lead to strong stratospheric warming which would limit the usefulness of our results. Nevertheless, our calculations indicate that at $10^{-4}$ of \chem{CH_4} the combined thermal and solar radiative forcings are 7.6\,\unit{W\,m^{-2}} (2\,\unit{bar} of $\mathrm{N_2}$), 7.2\,\unit{W\,m^{-2}} (1\,\unit{bar} of $\mathrm{N_2}$), and 6.2\,\unit{W\,m^{-2}} (0.5\,\unit{bar} of $\mathrm{N_2}$) and the thermal radiative forcings are 9.8\,\unit{W\,m^{-2}} (2\,\unit{bar} of $\mathrm{N_2}$), 8.6\,\unit{W\,m^{-2}} (1\,\unit{bar} of $\mathrm{N_2}$), and 6.8\,\unit{W\,m^{-2}} (0.5\,\unit{bar} of $\mathrm{N_2}$). Therefore, excluding the effects of overlap \citep[which are minimal,][]{Byrne-2014}, the combined thermal and solar radiative forcing due to $10^{-3}$ of \chem{CO_2} and $10^{-4}$ of \chem{CH_4} are 42.6\,\unit{W\,m^{-2}} (2\,\unit{bar} of $\mathrm{N_2}$), 33.2\,\unit{W\,m^{-2}} (1\,\unit{bar} of $\mathrm{N_2}$) and 21.2\,\unit{W\,m^{-2}} (0.5\,\unit{bar} of $\mathrm{N_2}$), significantly short of the forcings needed to sustain modern surface temperatures. It should be noted that strong solar absorption makes the precise radiative forcing highly sensitive to the position of the tropopause because this is the altitude at which most of the shortwave absorption is occurring. Therefore, small changes in the position of the tropopause result in large changes in the shortwave forcing. The surface temperature response to this forcing is less straight forward and the linearity between forcing and surface temperature change may breakdown \citet{Hansen-2005}.
 
  These calculations do not consider the products of atmospheric chemistry. Numerous studies have found that high \chem{CH_4}:\chem{CO_2} ratios lead to the formation of organic haze in low \chem{O_2} atmospheres which exerts an anti-greenhouse effect \citep{Kasting-1983,Zahnle-1986,Pavlov-2000,Haqq-Misra-2008}. Organic haze has been predicted by photochemical modelling at \chem{CH_4}:\chem{CO_2} ratios larger than 1 \citep{Zahnle-1986}, and laboratory experiments have found that organic haze could form at \chem{CH_4}:\chem{CO_2} ratios as low as 0.2--0.3 \citep{Trainer-2004,Trainer-2006}.

       \begin{figure*}
\includegraphics[height=70mm]{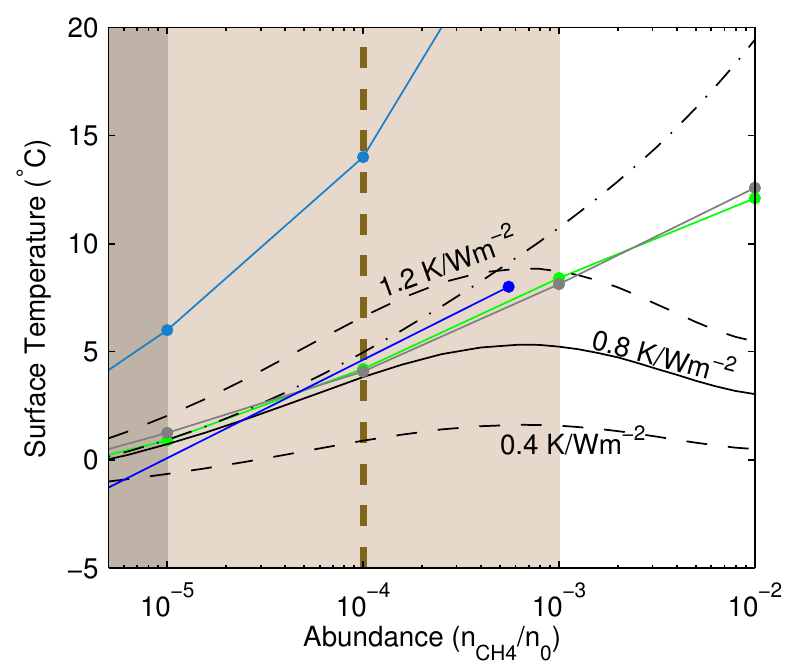}
 \caption{\textit{Surface temperature as a~function of \chem{CH_4} abundance for $\mathit{0.8 \, S_0}$}. Temperatures are calculated from radiative forcings assuming a~surface temperature of 271~K for 0~ppmv of \chem{CH_4} and climate sensitivity parameters of 0.4~$\mathrm{K/Wm^{-2}}$ (dashed black), 0.8~$\mathrm{K/Wm^{-2}}$ (solid black) and 1.2~$\mathrm{K/Wm^{-2}}$ (dashed black) and a~background \chem{CO_2} abundance of $10^{-2}$. Dashed-Dotted black line shows the longwave radiative forcing. The results of \citet{Wolf-2013} (blue), \citet{Haqq-Misra-2008} (green), \citet{Pavlov-2000} (turquoise), and \citet{Kiehl-1987} (grey) are also plotted. Temperatures for \citet{Kiehl-1987} are found from radiative forcings assuming a~climate sensitivity parameter of 0.81 $\mathrm{K/Wm^{-2}}$ and a~surface temperature of 271~K for 0~ppmv of \chem{CH_4}.}
 \label{fig:CH4Temp}
 \end{figure*}
 
As with \chem{CO_2}, we compare our \chem{CH_4} radiative forcings to values given in literature (Fig.~\ref{fig:CH4Temp}). Temperatures are calculated from radiative forcings assuming a~surface temperature of 271~K for an abundance of 0 and climate sensitivity parameters of 0.4~$\mathrm{K/Wm^{-2}}$, 0.8~$\mathrm{K/Wm^{-2}}$ and 1.2~$\mathrm{K/Wm^{-2}}$, and a~background \chem{CO_2} abundance of $10^{-2}$. Due to absorption of shortwave radiation, our calculated surface temperatures decrease for abundances above $10^{-3}$. Assuming a~linear relationship between forcing and climate response is likely a~poor assumption given strong atmospheric solar absorption. Results from \citet{Pavlov-2000} are included even though they are known to be erroneous as an illustration of the utility of these comparisons. All other studies give similar surface temperatures. However, these studies lack the strong solar absorption from the HITRAN 2012 database. \citet{Haqq-Misra-2008} shortwave radiative transfer is parametrized from data which pre-dates HITRAN 2000 and \citet{Wolf-2013} only include \chem{CH_4} absorption below 4650~$\mathrm{cm^{-1}}$ where changes to the spectra are not significant.

\subsection{Trace gases} \label{section:TraceGases}

The chemical cycles of several other greenhouse gases have been studied in the Archean. It has been hypothesized that higher atmospheric abundances could have been sustained making these gases important for the planetary energy budget. High abundances of \chem{NH_3} \citep{Sagan-1972}, \chem{C_2H_6} \citep{Haqq-Misra-2008}, \chem{N_2O} \citep{Buick-2007}, and OCS \citep{Ueno-2009} have all been proposed in the Archean. There are many other greenhouse gases in the HITRAN database that have not been studied, whether these gases could have been sustained at radiatively important abundances is beyond the scope of this paper. We review the sources, concentrations, and lifetimes from the strongest gases in table \ref{tab:GasTable} (modern Earth). We also provide a~review of the relevant literature on these gases in the Archean. Here we quantify the warming these gases could have provided in the Archean, motivated by future proposals of these as warming agents.

\subsubsection{Radiative forcings}

     \begin{figure*}
\includegraphics[height=100mm]{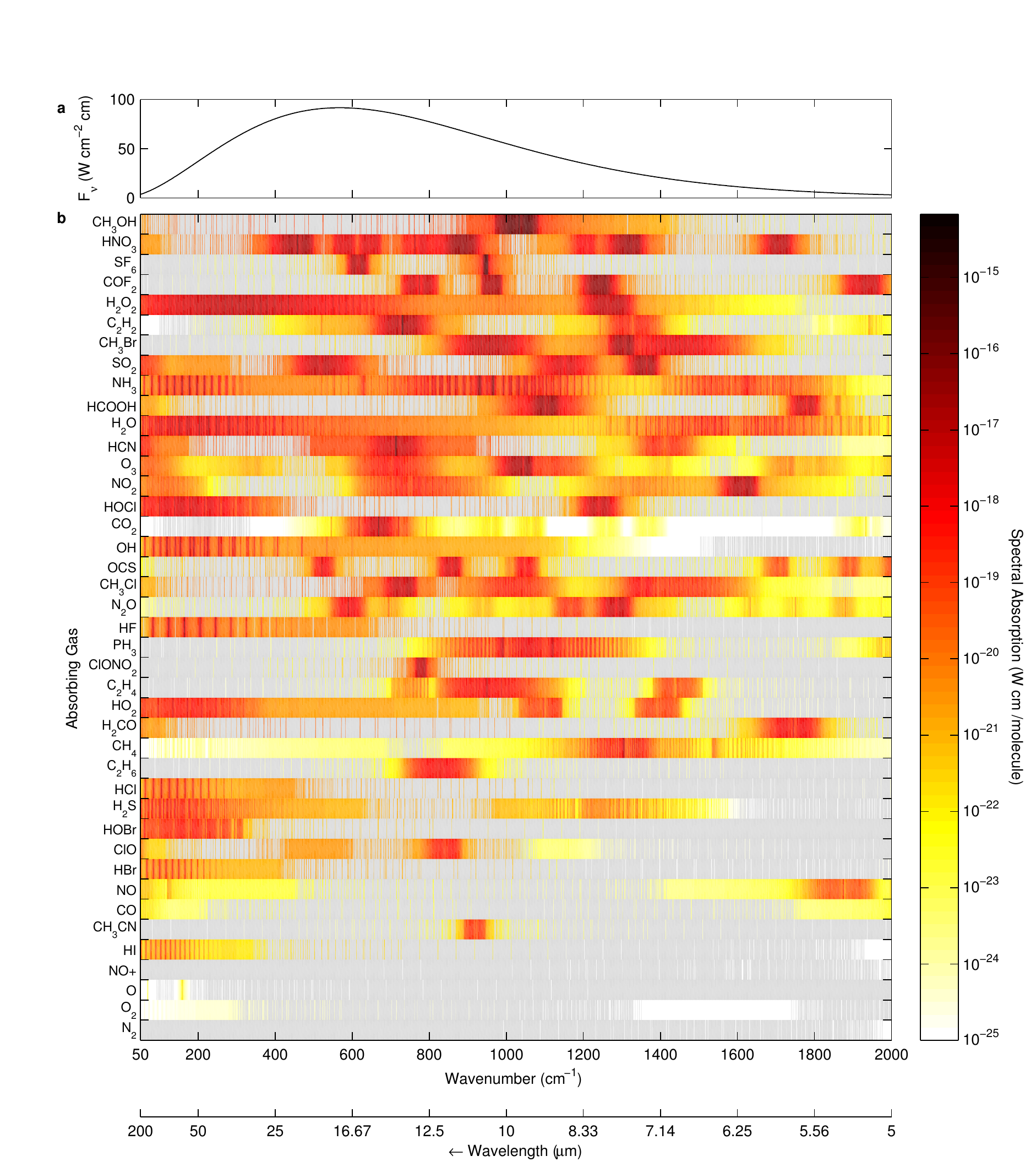}
       \caption{\textit{Spectral absorption of blackbody emissions.}
         \textbf{(a)} Emission intensity from a~blackbody of
         289\,\unit{K}. \textbf{(b)} Product of emission intensity and
         absorption cross-sections for gases from the HITRAN 2012
         database. Gases are ordered by decreasing spectrum integrated
         absorption strength from top to bottom. Grey indicates
         wavenumbers where no absorption data is available. Absorption
         coefficients were calculated at 500\,\unit{hPa} and
         260\,\unit{K}.}
 \label{fig:absorption}
 \end{figure*}
 
We produce a~first order estimate of the relative absorption strength of the HITRAN gases by taking the product of the irradiance produced by a~blackbody of 289~K and the absorption cross-sections to get the absorption per molecule of a~gas when saturated with radiation (Fig.~\ref{fig:absorption}). Using this metric, \chem{H_2O} ranks as the 11$\mathrm{^{th}}$ strongest greenhouse gas, and \chem{CO_2} and \chem{CH_4} rank 16$\mathrm{^{th}}$ and 27$\mathrm{^{th}}$ respectively. This demonstrates that many of the HITRAN gases are strong greenhouse gases and that it is conceivable that low abundances of these gases could have a~significant effect on the energy budget.

   \begin{figure*}
\includegraphics[height=90mm]{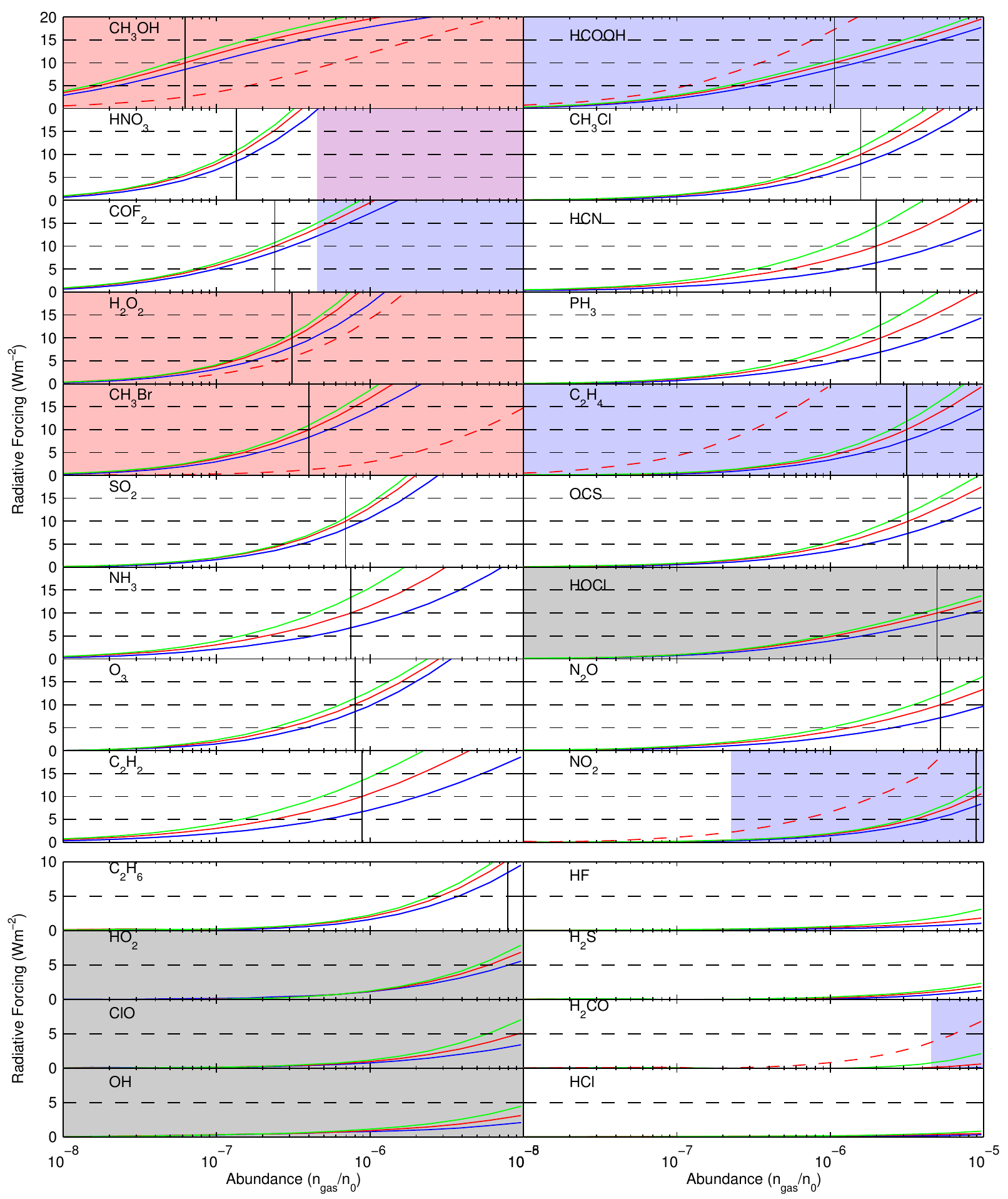}
     \caption{\textit{Trace gas radiative forcings.} All sky
       radiative forcings for potential early Earth trace gases,
       colors are as in Fig.~2. Gases are ordered by the
       concentration required to get a~radiative forcing of
       10\,\unit{W\,m^{-2}}. Shading indicates concentrations where
       computed cross-sections from HITRAN data were in poor agreement
       with the PNNL data. The colors indicate areas where HITRAN data
       underestimates (blue), overestimates (red), or both at
       different frequencies (purple). Grey shading indicates where no
       PNNL data was available. Vertical black lines show the
       concentration at which the radiative forcing is
       10\,\unit{W\,m^{-2}} for a~1\,\unit{bar} atmosphere. Dotted red
       lines give rough estimates of the radiative forcing accounting
       for incorrect spectral data. Concentrations are scaled to an
       atmosphere with 1\,\unit{bar} of \chem{N_2}.}
 \label{fig:forcing1}
 \end{figure*}
 
We calculate the radiative forcings for the HITRAN gases which produce forcings greater than 1\,\unit{W\,m^{-2}} over the abundances of $10^{-8}$ to $10^{-5}$ (Fig.~\ref{fig:forcing1}), assuming the gases are well-mixed. The radiative forcings are calculated in an atmosphere which contains only \chem{H_2O} and $\mathrm{N_2}$. Many of the gases reach forcings greater than 10 $\mathrm{Wm^{-2}}$ at abundances less than $10^{-6}$.

We give rough estimates of the expected radiative forcings assuming the PNNL cross-sections are correct for gases for which the HITRAN and PNNL cross-sections disagree. We have made approximate corrections to the forcings as follows. For some gases the shape of the absorption cross-sections were the same but the magnitude was offset. For these gases, we adjust the abundances required for a~given forcing, this was done for \chem{CH_3OH} (x10), \chem{CH_3Br} (x13), \chem{C_2H_4} (x0.1), and  \chem{H_2O_2} (x2). Missing spectra was compensated for by adding the radiative forcings from other gases that had similar spectra. For \chem{CH_3OH} the \chem{C_2H_4} forcings were added. For \chem{HCOOH} we added the HCN forcing. For \chem{NO_2} we added the HOCl forcing. For $\mathrm{H_2CO}$ we added the \chem{PH_3} forcing for a~given abundance scaled up an order of magnitude.

There are significant differences in radiative forcing due to different $\mathrm{N_2}$ inventories. The differences in radiative forcing due to differences in atmospheric pressure varies from gas to gas. Generally, the differences in forcing due to differences in atmospheric structure are similar, but the differences due to pressure broadening are more variable. Broadening is most effective for gases which have broad absorption features with highly variable cross-sections because the broadening of the lines covers areas with weak absorption. Such gases include \chem{NH_3}, \chem{HCN}, \chem{C_2H_2} and \chem{PH_3}. At $5 {\times} 10^{-6}$, 55-60 percent of the difference in radiative forcing between atmospheres can be attributed to pressure broadening for these gases. Where as, \chem{NO_2} and HOCl which have strong but narrow absorption features show the least difference in forcing due to pressure broadening (20-23\,{\%}).

\subsubsection{Overlap}

Here we examine the reduction in radiative forcing due to overlap for gases which reach radiative forcings of 10\,\unit{W\,m^{-2}} at abundances less than $10^{-5}$. The abundances of \chem{CO_2} and \chem{CH_4} are expected to be quite high in the Archean. Trace gases which have absorption bands coincident with the absorption bands of \chem{CO_2} and \chem{CH_4} will be much less effective at warming the Archean atmosphere.

 \begin{figure*}
\includegraphics[width=120mm]{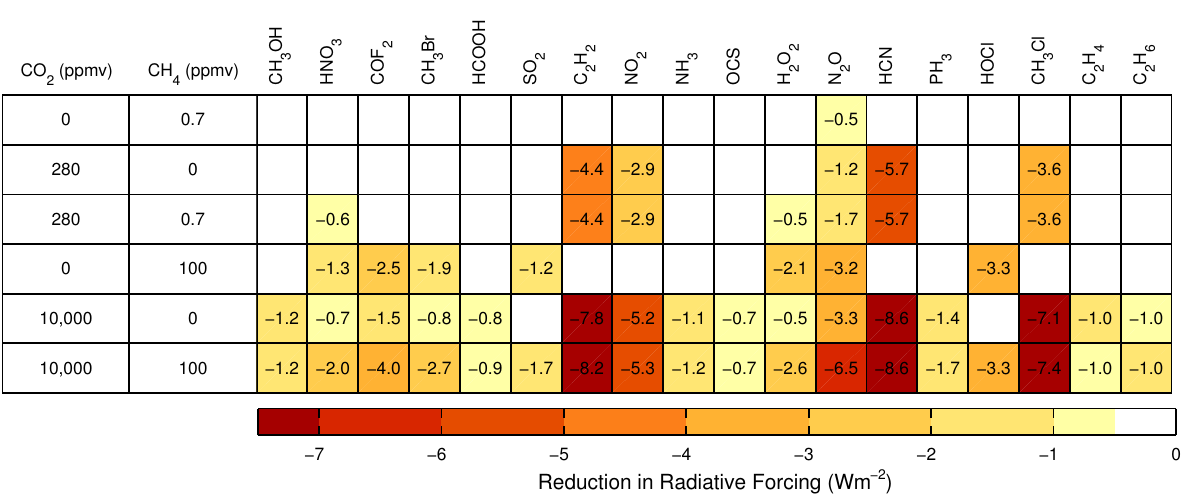}
\caption{\textit{Overlap with} \chem{CO_2} \textit{and} \chem{CH_4}.
  Reduction in radiative forcing due to overlapping absorption. Trace
  gas concentrations are held at the concentrations which gives
  a~10\,\unit{W\,m^{-2}} radiative forcing for an atmosphere with
  1\,\unit{bar} of \chem{N_2}.}
 \label{fig:overlap1}
 \end{figure*}

   \begin{figure*}
\includegraphics[width=120mm]{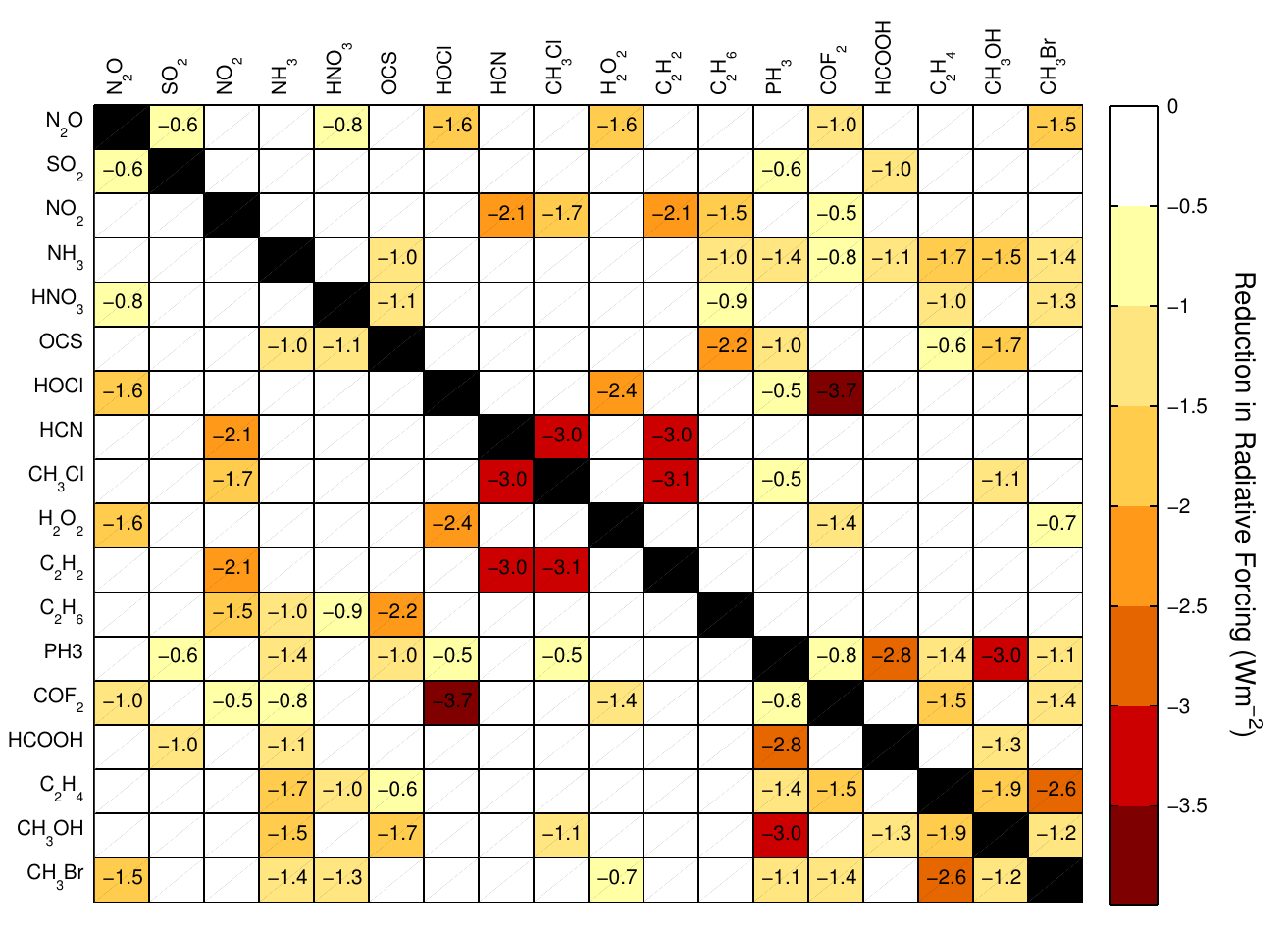}
     \caption{\textit{Trace gas overlap.} Reduction in radiative
       forcing due to overlapping absorption. Gas concentrations are
       held at concentrations which give a~10\,\unit{W\,m^{-2}}
       radiative forcing for an atmosphere with 1\,\unit{bar} of
       \chem{N_2}.}
 \label{fig:overlap2}
 \end{figure*}

We examine the effect of overlap on radiative forcing by looking at several cases with varying abundances of \chem{CO_2}, \chem{CH_4} (Fig.~\ref{fig:overlap1}), and other trace gases (Fig.~\ref{fig:overlap2}). The magnitude of overlap can vary substantially between the gases in question. For the majority of gases overlap with \chem{CO_2} is the largest. The reduction in forcing is generally between 10--30\,{\%} but can be as high as 86\,{\%}. The reduction in forcing are largest for \chem{HCN} (86\,{\%}), \chem{C_2H_2} (78\,{\%}), \chem{CH_3Cl} (71\,{\%}), \chem{NO_2} (52\,{\%}), and \chem{N_2O} (33\,{\%}) all with $10^{-2}$ of \chem{CO_2}. All of these gases have significant absorption bands in the 550-850~$\mathrm{cm^{-1}}$ wavenumber region where \chem{CO_2} absorbs the strongest. Of particular interest is \chem{N_2O} which has previously been proposed to have built up to significant abundances on the early Earth \citep{Buick-2007}. \chem{C_2H_2} could also have been produced by a~hypothetical early Earth haze, although previous studies have found that it would not build up to radiatively important abundances \citep{Haqq-Misra-2008}.

The reduction in forcing due to \chem{CH_4} is generally less than 20\,{\%} but can be as high as 37\,{\%}. The reductions in forcing are largest for HOCl (33\,{\%}), \chem{N_2O} (32\,{\%}), \chem{COF_2} (25\,{\%}), and \chem{H_2O_2} (21\,{\%}) all with $10^{-4}$ of \chem{CH_4}. All of which have absorption bands in 1200--1350~$\mathrm{cm^{-1}}$. As with \chem{CO_2}, the radiative forcing from \chem{N_2O} is significantly reduced due to overlap with \chem{CH_4}, suggesting that \chem{N_2O} is not a~good candidate to produce significant warming on early Earth except at very high abundances.

We calculate the reduction in radiative forcing due to overlap between trace gases (Fig.~\ref{fig:overlap2}). There is a~large amount of overlap between \chem{C_2H_2}, \chem{CH_3Cl} and \chem{HCN} resulting in a~reduction in radiative forcing of $\approx30\,{\%}$ . All three gases have their strongest absorption bands in the region 700--850~$\mathrm{cm^{-1}}$ and have a~secondary absorption band in the region 1250--1500~$\mathrm{cm^{-1}}$ which are on the edges of the water vapour window. All three gases have significant overlap with \chem{CO_2}, for an atmosphere with $10^{-2}$ of \chem{CO_2} the reductions in forcing are $>70\,{\%}$. 

Other traces gases with significant overlap are \chem{COF_2} and HOCl (37\,{\%}) due to coincident absorption bands at $\approx 1250 \mathrm{cm^{-1}}$, and \chem{CH_3OH} and \chem{PH_3} (30\,{\%}) due to coincident absorption around $~\approx 1000 \mathrm{cm^{-1}}$.

Other background absorption could have been present which would have had overlapping absorption with these gases. \citet{Wordsworth-2013} have proposed that elevated $\mathrm{N_2}$ and $\mathrm{H_2}$ levels may have been present in the Archean which would have resulted in significant $\mathrm{N_2}$--$\mathrm{H_2}$ CIA across much of the infrared spectrum including the water vapour window. Overlap between $\mathrm{N_2}$--$\mathrm{H_2}$ absorption and the gases examined here would likely be significant as many of these gases have significant absorption in the water vapour window. However, performing these overlap calculations is beyond the scope of this work.

\subsubsection{Comparison between our results and previous calculations}

    \begin{figure*}
\includegraphics[height=100mm]{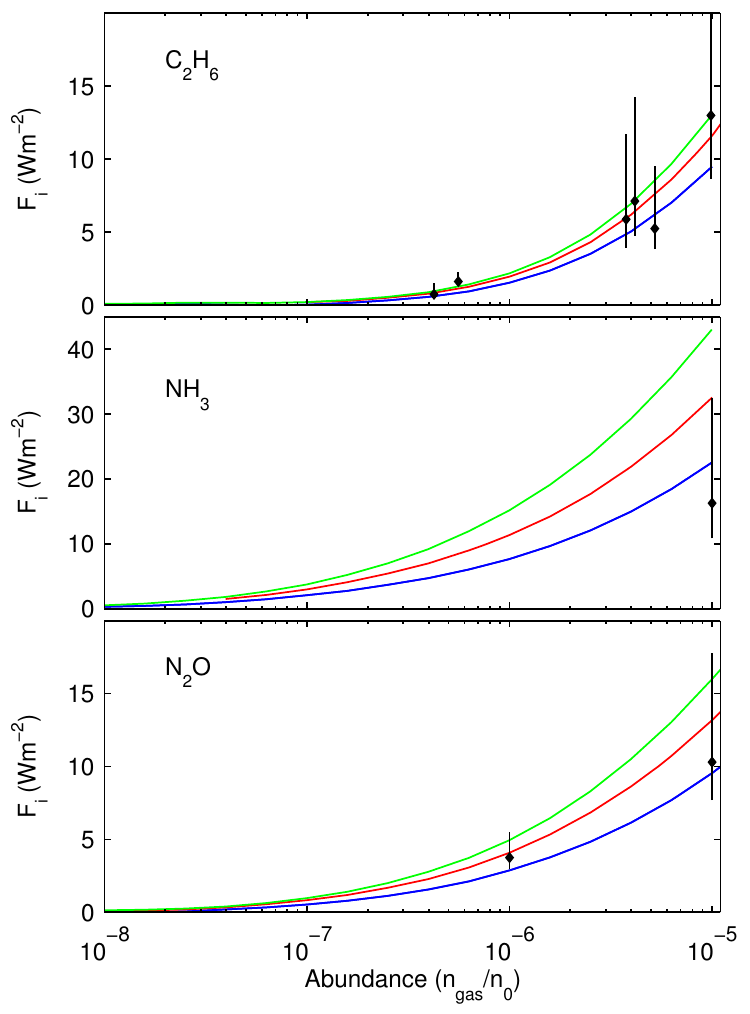}
      \caption{\textit{Calculated Radiative forcings and inferred
          radiative forcings from literature.} Literature radiative
        forcings are inferred from temperature changes reported by
        \citet{Haqq-Misra-2008} (\chem{C_2H_6}), \citet{Kuhn-1979}
        (\chem{NH_3}) and \citet{Roberson-2011}
        (\chem{N_2O}). Radiative forcings are calculated assuming
        a~range of climate sensitivity parameters of
        0.4~to 1.2\,K(Wm$^{-2}$)$^{-1}$ with
        a~best guess of 0.8\,K(Wm$^{-2}$)$^{-1}$.}
 \label{fig:forcingComparison}
 \end{figure*}

We compare inferred radiative forcings from prior work to ours using the same method as for \chem{CO_2} and \chem{CH_4} (Fig.~\ref{fig:forcingComparison}).

Inferred \chem{C_2H_6} radiative forcings from \citet{Haqq-Misra-2008} for a~1\,\unit{bar} atmosphere agree well with our results. Inferred \chem{NH_3} from \citet{Kuhn-1979} for a~0.78\,\unit{bar} atmosphere agree well with our results, although, our results suggest that the results of \citet{Kuhn-1979} are on the lower end of possible temperature changes. Inferred \chem{N_2O} from \citet{Roberson-2011} for a~1\,\unit{bar} atmosphere agree well with our results. \citet{Roberson-2011} perform radiative forcing calculations with \chem{CO_2} and \chem{CH_4} abundances of 320~ppmv and 1.6~ppmv. Due to overlap this forcing is likely reduced by $\approx$50\,{\%} with early Earth \chem{CO_2} and \chem{CH_4} abundances of $10^{-2}$ and $10^{-4}$. \citet{Ueno-2009} give a~rough estimate of the radiative forcing due to 10~ppmv of OCS to be 60\,\unit{W\,m^{-2}}. In this work we find the forcing to be much less than this ($\approx\,20\,\mathrm{Wm^{-2}}$).

\conclusions

Using the SMART radiative transfer model and HITRAN line data, we have calculated radiative forcings for \chem{CO_2}, \chem{CH_4} and 26 other HITRAN greenhouse gases on a~hypothetical early Earth atmosphere. These forcings are available at several background pressures and we account for overlap between gases. We recommend the forcings provided here be used both as a~first reference for which gases are likely good greenhouse gases, and as a~standard set of calculations for validation of radiative forcing calculations for the Archean. Model output are available as supplementary material for this purpose. Many of these gases can produce significant radiative forcings at low abundances. Whether any of these gases could have been sustained at radiatively important abundances during the Archean requires study with geochemical and atmospheric chemistry models.

Comparing our calculated forcings with previous work, we find that \chem{CO_2} radiative forcings are consistent, but find a~stronger shortwave absorption by \chem{CH_4} than previously recorded. This is primarily due to updates to the HITRAN databse at wavenumbers less than 11,502~$\mathrm{cm^{-1}}$. This new result suggests an upper limit to the warming \chem{CH_4} could have provided of about 10\,\unit{W\,m^{-2}} at $5{\times}10^{-4}$--$10^{-3}$. Amongst the trace gases, we find that the forcing from \chem{N_2O} was likely overestimated \citet{Roberson-2011} due to underestimated overlap with \chem{CO_2} and \chem{CH_4}, and that the radiative forcing from OCS was greatly overestimated by \citet{Ueno-2009}.

\appendix
\section{Sensitivity}
\label{section:appendix}

\begin{figure*}
\includegraphics[width=120mm]{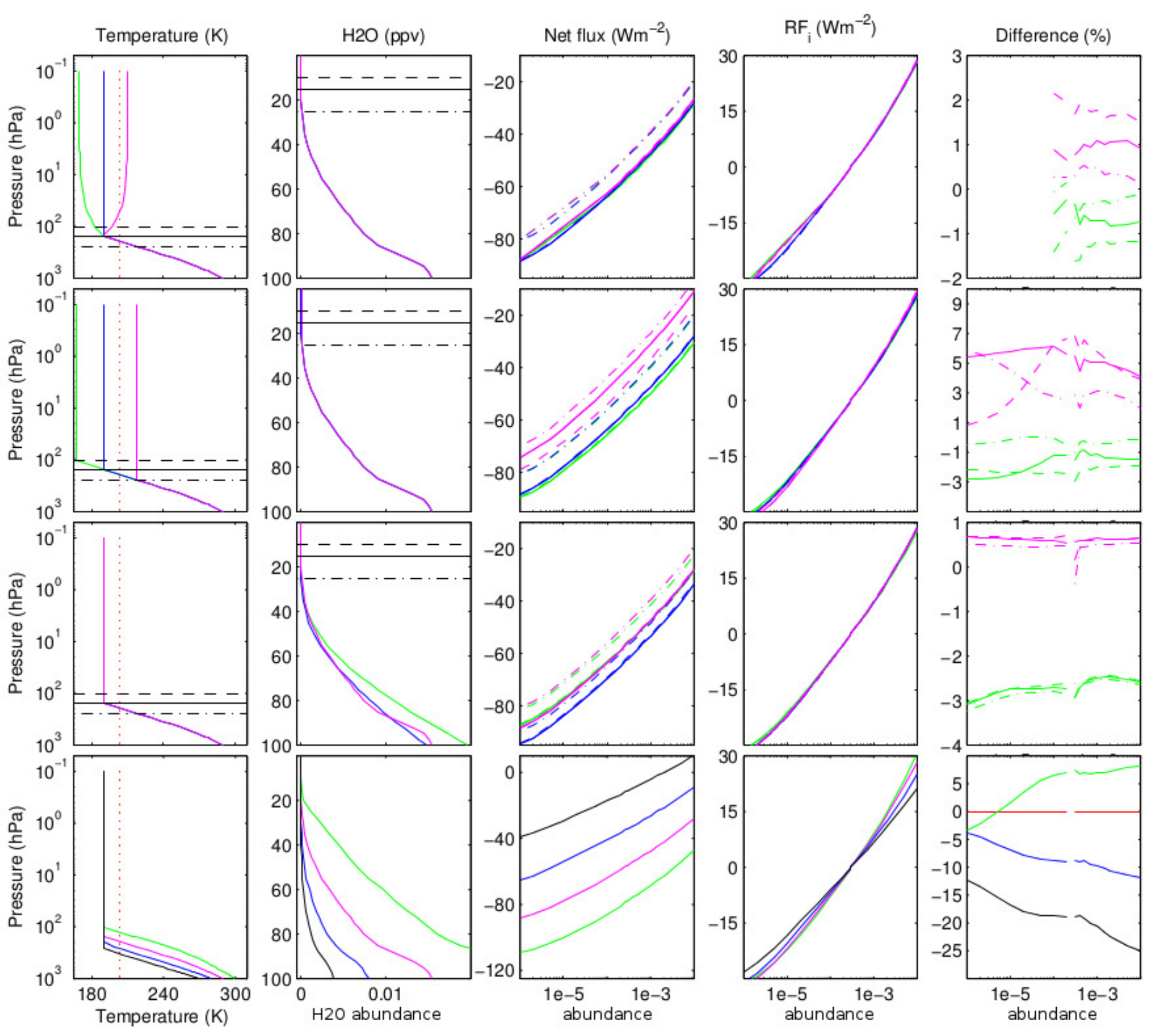}
\caption{\textit{Sensitivity Study.} Columns from left to right:
  Temperature structure, \chem{H_2O} structure, Net flux of radiation
  at tropopause, radiative forcing, and percentage difference in
  radiative forcing. Solid, dashed and dashed-dotted curves represent
  different tropopause positions. Vertical dotted red line shows the
  atmospheric skin temperature (203\,\unit{K}).}
 \label{fig:sensitivity}
    \end{figure*}

Figure \ref{fig:sensitivity} shows the fluxes, radiative forcings and
percentage difference in radiative forcings for various possible GAM
temperature and water vapour profiles for the early Earth. The effect
on radiative forcing calculations from varying the stratospheric
\mbox{temperature} from 170~to 210\,\unit{K} while the tropopause
temperature is kept constant are very small ($<3\,{\,{\%}}$). Varying the
tropopause temperature between 170~and 210\,\unit{K} results
in larger differences in radiative forcing ($<10\,{\,{\%}}$). Changing the
relative humidity effects radiative forcing by less than
5\,{\,{\%}}. Radiative forcing calculations are sensitive to surface
temperature. Increasing or decreasing a~290\,K surface temperature by
10\,K results in differences in radiative
forcing of $\le 12$\,{\,{\%}}. However, the difference in forcing between
270~and 290\,\unit{K} is much larger (12--25\,{\,{\%}}).

\Supplementary{pdf}

\begin{acknowledgements}
We thank Ty Robinson for help with SMART and discussions of the theory behind it. Financial support was received from the Natural Sciences and Engineering Research Council of Canada (NSERC) CREATE Training Program in Interdisciplinary Climate Science at the University of Victoria (UVic); a~University of Victoria graduate fellowship to B.B. and NSERC Discovery grant to C.G. This research has been enabled by the use of computing resources provided by WestGrid and Compute/Calcul Canada. We would also like to thank Andrew MacDougall for helpful comments on an earlier draft of the manuscript and Kevin Zahnle for sharing is insights and reviewing the information in table 1. We thank Jim Kasting and an anonymous reviewer for comments which improved the manuscript.
\end{acknowledgements}

\appendix
\onecolumn

%% \begin{table}[t]
%%\caption{\textit{Global Atmospheric Budget of 20 strongest HITRAN gases.} The sources, sinks, lifetimes and concentrations of these gases are given for the modern atmosphere. Relevant literature for the Archean is reviewed.}\label{tab:GasTable}
%{\scriptsize
%\begin{threeparttable}
%{
\begin{longtable}{ p{1cm}  p{2.5cm}  p{4cm}  p{4cm} }
\caption{\textit{Global Atmospheric Budget of 20 strongest HITRAN gases.} The sources, sinks, lifetimes and concentrations of these gases are given for the modern atmosphere. Relevant literature for the Archean is reviewed.} \label{tab:GasTable} \\
\hline
 Gas & \underline{Modern} \newline F\,=\,Flux $\mathrm{(moles/yr)}$  \newline x\,=\,Abundance ($n_{gas}/n_0$) \newline $\tau$\,=\,Atmospheric lifetime & Modern Natural Sources and Sinks & Notes for Archean \\ %\tnote{a}
%%& Flux $\mathrm{\left(moles/yr \right)}$ \newline / \newline abundance (ppv) \newline / \newline Atmospheric lifetime & Major Natural Sources and Sinks & \\
\hline
\chem{CH_3OH} \newline (methanol) & F\,=\,$\mathrm{2.3 {\times} 10^{12}}$--$\mathrm{1.5 {\times} 10^{13}}$ $[1,2]$ \newline x\,=\,$\mathrm{10^{-9}}$--$\mathrm{10^{-8}}$ (boundary layer) \newline $\mathrm{10^{-10}}$--$\mathrm{10^{-9}}$ (free troposphere)$[1]$ \newline $\tau$\,=\,5--12\,days$[1]$ & Largest known sources are plant growth and decay (chemical and enzymatic demethylation of methoxy groups)$[1]$. The ocean biosphere may also be a~significant source$[2]$. The main sink is oxidation by OH$[1]$, although, deposition$[1]$ and possibly ocean uptake$[2]$ are important. & Photolysis of \chem{H_2O} in the presence of CO leads to the production of \chem{CH_3OH}\,$[3]$. Atmospheric modelling suggests this could have produced surface abundances of $\approx 2 {\times} 10^{-13}$ in the Archean$[4]$. Methanotrophs produce methanol as an intermediate product which is then oxidized via formaldehyde and formate to carbon dioxide, however, it has been found that high levels of \chem{CO_2} can partially inhibit methanol oxidation$[5]$. \\
\hline
%% ----
\chem{HNO_3} \newline (nitric acid)  & F\,=\,unknown? \newline x\,=\,variable \newline $\tau$\,=\,2.6~hours$[6]$ & Major sources are denitrification and lightning$[6]$. Lightning strikes produced NO from \chem{O_2} and $\mathrm{N_2}$, which is oxidized to \chem{NO_2} and then to \chem{HNO_3}. Major sinks are deposition and dilution$[6]$. & Early Earth volcanism may have produced $\approx10^9$--$10^{10}$~mol/yr of fixed nitrogen compared to $\approx10^9$~mol/year today$[7]$. Without atmospheric \chem{O_2}, the nitrogen fixation efficiency of lightning is significantly reduced compared to today$[8]$. Although, it has also been suggested that nitrogen fixation by volcanic lightning could have produced $\mathrm{3.3 {\times} 10^{10}}$--$\mathrm{3.3 {\times} 10^{11}}$~mol/yr of NO (4~Gyr Ago)$[9]$. However, in reduced atmospheres NO reacts to form HNO (K. Zahnle, private communication). \\
\hline
%% ----
 \chem{COF_2} \newline (carbonyl fluoride)  & F\,=\,unknown? \newline x\,=\,$\mathrm{10^{-10}}$--$\mathrm{10^{-9}}$\,(mid atmosphere)$[10]$ $\mathrm{<10^{-10}}$\,(troposphere)$[10]$ \newline $\tau$\,=\,$\approx 3.8$\,years$[10]$ & Produced in the stratosphere from photolysis of chlorofluorocarbons$[11]$. & The vast majority of atmospheric fluorine emissions come in the form of man-made emissions$[10]$ \\
\hline
% % ----
 \chem{H_2O_2} \newline (hydrogen peroxide)  & F\,=\,unknown \newline x\,=\,$\mathrm{10^{-10}}$--$\mathrm{10^{-9}}$\,$[12]$ \newline $\tau$\,=\,a few days$[12]$ & No significant direct emissions have been found. Formed by bimolecular and termolecular recombination of $\mathrm{HO_2}$ and radicals during the day time$[13]$. Sinks include washout, dry deposition, photolysis and reactions with OH$[13]$. & \chem{H_2O} photolysis produces \chem{H_2O_2}. It has been estimated that this process could produce tropospheric abundances of $\leq1.7 {\times} 10^{-15}$\,$[14]$. During intense glaciation higher abundances of $\approx10^{-10}$ may have been sustained by \chem{H_2O} photolysis$[15]$  \\
\hline
% % ----
 \chem{CH_3Br} \newline (methyl bromide), \newline (bromomethane) & F\,=\,$\mathrm{1 {\times} 10^9}$ $[16]$ \newline x\,=\,$\approx 5 {\times} 10^{-12}$ (pre-industrial)$[17]$ \newline $\tau$\,=\,0.8~years$[16]$  & Major natural sources are oceans, freshwater wetlands and coastal salt marshes, though other sources may be important$[16]$. Major sinks include oceans, oxidation by OH, photolysis, and soil microbial uptake$[16]$. & Early Earth volcanism may have produced greater fluxes of bromine than today ($10^8$-$10^{9}$~mol/year of Br$[17]$). Siberian trap volcanism may have produced large \chem{CH_3Br} fluxes ($\mathrm{2 {\times} 10^{10}}$--$\mathrm{5.3 {\times} 10^{10}}$ mol/year)$[18]$ \\
\hline
% % ----
 \chem{SO_2} \newline (sulphur dioxide)  & F\,=\,$\mathrm{1.2 {\times} 10^{12}}$ $[19]$ \newline x\,=\,variable \newline $\tau$\,=\,8--78 hours$[20]$ & Major sources are dimethyl sulphide (DMS) related (75\,{\%}) and from volcanoes (25\,{\%})$[19]$. Major sinks are OH, dry deposition aerosol scavenging, and conversion to \chem{H_2SO_4}. & Emergence of continents and subaerial volcanism may have resulted in much increased \chem{SO_2} emissions in the late Archean compared to the early Archean$[21]$. Estimates of \chem{SO_2} released by Archean volcanism are as high as $\approx\mathrm{3 {\times} 10^{12}}$~mol/year$[22]$. Modelling results suggest that $\approx10^{-10}$ of \chem{SO_2} could have existed in the Archean atmosphere if volcanism emitted $\approx\mathrm{1 {\times} 10^{12}}$~mol/year$[23]$. It has been suggested that \chem{SO_2} abundances must have remained relatively low during the Earth's history as high abundances would have inhibited calcium carbonate precipitation$[24]$.\\
\hline
% % ----
\chem{NH_3} \newline (ammonia)  & F\,=\,$\mathrm{1.3 {\times} 10^{12}}$\,$[6]$, $\mathrm{5.4 {\times} 10^{12}}$\,$[25]$ \newline x\,=\,$6.7{\times} 10^{-10}$ (combined \chem{NH_3} and \chem{NH_4})$[6]$ \newline $\tau$\,=\,1--5\,days$[26]$ & Sources are through biological fixation$[6]$. Specific sources are domesticated animals ($\approx43\,{\%}$), sea surface ($\approx17\,{\%}$), undisturbed soils ($\approx13\,{\%}$), fertilizers ($\approx12\,{\%}$), biomass burning ($\approx7\,{\%}$), and others ($\approx8\,{\%}$)$[25]$. Deposition is the largest sink$[6]$. Specific sinks are wet deposition on land ($\approx53\,{\%}$), wet deposition on sea surface ($\approx28\,{\%}$), dry deposition on land ($\approx18\,{\%}$) and reaction with OH ($\approx2\,{\%}$)$[25]$. & \chem{NH_3} could have been generated by hydrolysis of HCN in the stratosphere$[27]$. Although even with UV shielding by an organic haze abundances greater than seem $10^{-9}$ unlikely$[28]$. Large biological flux would be possible in absence of nitrification. See section~\ref{sec:Int} \\
\hline
%% ----
 \chem{O_3} \newline (ozone) & F\,=\,not emitted directly \newline x\,=\,$\mathrm{10^{-9}}$--$\mathrm{10^{-8}}$\,(troposphere)$[29]$ $\mathrm{\approx10^{-5}}$\,(stratosphere)$[30]$  \newline $\tau$\,=\,22~days (troposphere) & Chapman cycle: stratospheric \chem{O_3} is produced by \chem{O_2} photochemistry and is destroyed by reactions with atomic oxygen$[30]$. & With very low atmospheric \chem{O_2} abundances, \chem{O_3} abundances would be negligible$[31]$. \\
\hline
% % ----
 \chem{C_2H_2} \newline (acetylene)  & F\,=\,$\mathrm{1 {\times} 10^{11}}$ $[32]$--$\mathrm{2.5 {\times} 10^{11}}$ $[33]$ \newline x\,=\,$10^{-12}$--$10^{-9}$ $[33]$ \newline $\tau$\,=\,2\,weeks$[33]$ & Anthropogenic sources dominate: biomass burning, natural gas loss and biofuel consumption. Oxidation by OH is a~major sink. & Methane photochemistry may have sustained tropospheric abundances of $\approx \mathrm{2 {\times}10^{-14}}$ in the Archean (for \chem{CH_4} and \chem{CO_2} abundances of $10^{-3}$)\,$[34]$ \\
\hline
% % ----
 HCOOH \newline (formic acid) & F\,=\,$\mathrm{1.2 {\times} 10^{12}}$ $[35]$ \newline x\,=\,$\mathrm{3 {\times} 10^{-11}}$--$\mathrm{5 {\times} 10^{-9}}$ (rural)$[35]$ \newline $\tau$\,=\,1-10\,days$[35]$ & Major natural sources are bimass burning, soil, vegetation, as well as secondary production from gas-phase and aqueous photochemistry$[35]$. Very soluble and the major sink is thought to be deposition. Relatively long-lived with respect to oxidation by OH (25 days)$[35]$. & \\
\hline
% % ----
\chem{CH_3Cl} \newline (methyl chloride), \newline (chloromethane) & F\,=\,$\mathrm{7.2 {\times} 10^{10}}$-$\mathrm{9.2 {\times} 10^{10}}$ $[16]$ \newline x\,=\,$\mathrm{4.5 {\times} 10^{-10}}$\,$[36]$ \newline $\tau$\,=\,1\,year & Natural sources are biomass burning and oceans$[16]$. a~mechanism to produce \chem{CH_3Cl} through the photochemical reaction of dissolved organic matter in saline waters has also been reported$[37]$. Major sinks include oxidation by OH, reaction with chlorine radicals, uptake by oceans and soils, and loss to the stratosphere (for tropospheric \chem{CH_3Cl})$[16]$ & Early Earth volcanism may have produced greater fluxes of chlorine than today ($10^8$-$10^{10}$~mol/year of Cl$[17]$). Siberian trap volcanism may have produced large \chem{CH_3Cl} fluxes ($\mathrm{2 {\times} 10^{12}}$--$\mathrm{5.5 {\times} 10^{12}}$ mol/year)$[18]$. \chem{CH_3Cl} has been found to correlate very well with methane during the holocene (11--0\,kyr B.P.) and appears to correlate during the last glacial period (50--30\,kyr B.P.)$[38]$ \\
\hline
%% ----
 HCN \newline (hydrogen cyanide)  & F\,=\,$\mathrm{5.6 {\times} 10^{10}}$-$\mathrm{1.3 {\times} 10^{11}}$ $[39]$ \newline x\,=\,$\approx2{\times} 10^{-10}$\,$[40]$ \newline $\tau$\,=\,5.3\,months (troposphere)$[40]$ & Biomass Burning is a~major source$[39]$. Ocean uptake, oxidation by OH, photolysis, and scavenging by precipitation are major sinks. & Upper atmospheric chemistry between \chem{CH_4} photolysis products and atomic nitrogen may have produced elevated abundances in the Archean$[27,41]$. Perhaps sustaining tropospheric abundances of $\mathrm{10^{-8}}$--$\mathrm{10^{-7}}$ and higher abundances in the stratosphere$[41]$. HCN could be produced by lightning in a~reduced atmosphere with  $\mathrm{pCH_4}\geq\mathrm{pCO_2}$ (K. Zahnle, private communication). \\
\hline
% % ----
 \chem{PH_3} \newline (phosphine) & F\,=\,unknown \newline x\,=\,$<10^{-13}$\,(free troposphere)$[42]$ \newline $\tau$\,=\,28\,hours$[43]$ & Possible sources are Lightning and microbes$[44]$. The main sink is reaction with OH producing phosphate which returns to the surface in rain$[43]$. & Simulated lightning in a~methane model atmosphere has been shown to reduce phosphate to \chem{PH_3}\,$[44]$. Production of phosphine from posphate may also be produced via tectonic forces and processing of rocks$[45]$. \\
\hline
% % ----
 \chem{C_2H_4} \newline (Ethylene)  & F\,=\,$\mathrm{3 {\times} 10^{10}}$ \newline x\,=\,$\mathrm{10^{-11}}$--$\mathrm{10^{-8}}$\,$[46]$ \newline $\tau$\,=\,a few days$[46]$ & Major sources are plants and microorganisms. Destroyed by OH ($\approx90\,{\%}$) and \chem{O_3}($\approx10\,{\%}$)\,$[46]$. & Methane photochemistry may have sustained tropospheric abundances of $\approx \mathrm{2 {\times}10^{-13}}$ in the Archean (for \chem{CH_4} and \chem{CO_2} abundances of $10^{-3}$)\,$[34]$ \\
\hline
% % ----
 OCS \newline (Carbonyl sulfide)   & F\,=\,$\mathrm{1 {\times} 10^9}$-$\mathrm{3.3 {\times} 10^{10}}$ $[47]$ \newline x\,=\,$\mathrm{5 {\times} 10^{-10}}$\,$[47]$ \newline $\tau$\,=\,1.5-3~years$[47]$,4.3~years$[48]$ & Major sources are oceanic emissions, oxidation of oceanic $\mathrm{CS_2}$ and DMS, and biomass burning. Major sinks are uptake by vegetation and soils, and OH oxidation. OCS is oxidized in the stratosphere to form sulphate particles which influence the radiative budget$[47]$. & abundances of $\mathrm{5 {\times} 10^{-6}}$ have been suggested in the Archean$[49]$. However, other have found that abundances above $\mathrm{10^{-8}}$ appear unlikely due to rapid OCS photolysis$[50]$ $[23]$. See section~\ref{sec:Int} \\
\hline
% % ----
 HOCl \newline (hypochlorous acid)  & F\,=\,unknown? \newline x\,=\,$5{\times} 10^{-12}$--$1.7{\times} 10^{-10}$\,$[51]$ \newline $\tau$\,=\,days? & Cl can form HOCl via gas phase reactions$[21]$. & Early Earth volcanism may have produced greater fluxes of chlorine than today ($10^8$-$10^{10}$~mol/year of Cl$[17]$). \\
\hline
% % ----
 \chem{N_2O} \newline (nitrous oxide) & F\,=\,$\mathrm{8.6 {\times} 10^{11}}$ $[52]$ \newline x\,=\,$\mathrm{2.7 {\times} 10^{-7}}$\,$[53]$ \newline $\tau$\,=\,131\,years$[53]$  & Produced as an intermediate product of both nitrification and denitrification$[52]$. Primary natural sources are upland soils and riparian areas, oceans, estuaries, and rivers$[52]$. Destroyed by chemical reactions in the upper atmosphere. & It has been proposed that large amounts of \chem{N_2O} could have been produced in the Proterozoic due to bacterial denitrification in copper depleted water$[54]$. However, it has been shown that \chem{N_2O} would be rapidly photo-dissociated if \chem{O_2} levels were lower than 0.1 PAL$[55]$. \\
\hline
% % ----
 \chem{NO_2} \newline (nitrogen dioxide) & F\,=\,$\mathrm{7 {\times} 10^{11}}$-$\mathrm{3.6 {\times} 10^{12}}$ $[56]$ \newline x\,=\,$\mathrm{3 {\times} 10^{-10}}$\,(NO,\chem{NO_2}, and $\mathrm{NO_3}$)\,$[6]$ \newline $\tau$\,=\,27\,hours$[56]$  & Lightning, Biomass Burning, and soils are main sources$[57]$. Lightning strikes produced NO from \chem{O_2} and $\mathrm{N_2}$, which is oxidized to \chem{NO_2}. a~major sink is dry deposition$[6]$. & Early Earth volcanism may have produced $\approx10^9$--$10^{10}$~mol/yr of fixed nitrogen compared to $\approx10^9$~mol/year today$[7]$. Without \chem{O_2}, the nitrogen fixation efficiency of lightning is significantly reduced compared to today$[8]$. Although, it has also been suggested that nitrogen fixation by volcanic lightning could have produced $\mathrm{3.3 {\times} 10^{10}}$--$\mathrm{3.3 {\times} 10^{11}}$~mol/yr of NO (4~Gyr Ago)$[9]$. However, in reduced atmospheres NO reacts to form HNO (K. Zahnle, private communication). \\
\hline
% % ----
 \chem{C_2H_6} \newline (ethane) & F\,=\,$\mathrm{2.0 {\times} 10^{11}}$ $[32]$ \newline x\,=\,$10^{-10}$--$5{\times} 10^{-9}$ $[58]$ \newline $\tau$\,=\,2\,months$[58]$ & Anthropogenic sources dominate: biomass burning, natural gas loss and biofuel consumption. Oxidation by OH is a~major sink. & Methane photochemistry may have sustained tropospheric abundances of $\approx \mathrm{5 {\times}10^{-6}}$ in the Archean (for \chem{CH_4} and \chem{CO_2} abundances of $\mathrm{10^{-3}}$)$[34]$. See section~\ref{sec:Int} \\
\hline
%%HO2   & & & \\
%%ClO   & & & \\
%%OH    & & & \\
%%HF    & & & \\
%%H2S   & $\mathrm{7.7 {\times} 10^{10}}$-$\mathrm{1.2 {\times} 10^{11}}$ & & \\
%%H2CO  & & & \\
%%HCl   & & & \\
%\hline
%
\end{longtable}
%newpage
%%\begin{tablenotes}
\belowtable{
 $[1]$~\citet{Jacob-2005} and references within. 
 $[2]$~\citet{Millet-2008}
 $[3]$~\citet{Barnun-1983}
 $[4]$~\citet{Wen-1989}
 $[5]$~\citet{Xin-2004}
 $[6]$~\citet{Ussiri-2012}
 $[7]$~\citet{Mather-2004}
 $[8]$~\citet{Navarro-Gonzalez-2001}
 $[9]$~\citet{Navarro-Gonzalez-1998}
 $[10]$~\citet{Harrison-2014}
 $[11]$~\citet{Duchatelet-2009}
 $[12]$~\citet{Allen-2013}
 $[13]$~\citet{Hua-2008}
 $[14]$~\citet{Haqq-Misra-2011}
 $[15]$~\citet{Liang-2006}
 $[16]$~\citet{WMO-2010} 
 $[17]$~\citet{Martin-2007}
 $[18]$~\citet{Svensen-2009}
 $[19]$~\citet{Smith-2001}
 $[20]$~\citet{Lee-2011b}
 $[21]$~\citet{Gaillard-2011}
 $[22]$~\citet{Zahnle-2006}
 $[23]$~\citet{Zerkle-2012}
 $[24]$~\citet{Halevy-2009}
 $[25]$~\citet{Schlesinger-1992}
 $[26]$~\citet{Warneck-1988}
 $[27]$~\citet{Zahnle-1986}
 $[28]$~\citet{Pavlov-2001b}
 $[29]$~\citet{Fishman-1978}
 $[30]$~\citet{Johnston-1975}
 $[31]$~\citet{Kasting-1980}
 $[32]$~\citet{Abad-2011}
 $[33]$~\citet{Xiao-2007}
 $[34]$~\citet{Haqq-Misra-2008}
 $[35]$~\citet{Chebbi-1996}
 $[36]$~\citet{Williams-2007}
 $[37]$~\citet{Moore-2008}
 $[38]$~\citet{Verhulst-2013}
 $[39]$~\citet{Zeng-2012}
 $[40]$~\citet{Li-2003}
 $[41]$~\citet{Tian-2011}
 $[42]$~\citet{Glindemann-2003}
 $[43]$~\citet{Morton-2005}
 $[44]$~\citet{Glindemann-2004}
 $[45]$~\citet{Glindemann-2005}
 $[46]$~\citet{Sawada-1985}
 $[47]$~\citet{Kettle-2002} and \citet{Montzka-2007}
 $[48]$~\citet{Chin-1995}
 $[49]$~\citet{Ueno-2009}
 $[50]$~\citet{Domagal-Goldman-2011}
 $[51]$~\citet{Lawler-2011}
 $[52]$~\citet{EPA-2010}
 $[53]$~\citet{IPCC2013}
 $[54]$~\citet{Buick-2007}
 $[55]$~\citet{Roberson-2011} 
 $[56]$~\citet{Leue-2001}
 $[57]$~\citet{Lee-1997}
 $[58]$~\citet{Xiao-2008}
}
%%\end{tablenotes}
%\end{threeparttable} }
%%\end{table} 

\newpage

\appendixfigures

\end{document}